\documentclass[12pt,a4paper,twocolumn]{aastex61}
\pdfoutput=1 
\usepackage[USenglish]{babel}
\usepackage{amsmath,amstext}
\usepackage{graphicx}
\usepackage{xfrac}
\usepackage[T1]{fontenc}
\usepackage{apjfonts} 
\usepackage[figure,figure*]{hypcap}

\shorttitle{Cosmology with Dark Matter Halo Sparsity}
\shortauthors{P.S. Corasaniti et al.}

\begin{document}

\title{Probing Cosmology with Dark Matter Halo Sparsity Using X-ray Cluster Mass Measurements}

\author{P.S. Corasaniti}
\affiliation{LUTH, Observatoire de Paris, PSL Research University, CNRS, Universit\'e Paris Diderot, Sorbonne Paris Cit\'e, 5 Place Jules Janssen, 92195 Meudon, France}
\author{S. Ettori}
\affiliation{INAF, Osservatorio Astronomico di Bologna, via Piero Gobetti 93/3, I-40129 Bologna, Italy}
\affiliation{INFN, Sezione di Bologna, viale Berti Pichat 6/2, 40127 Bologna, Italy}
\author{Y. Rasera}
\affiliation{LUTH, Observatoire de Paris, PSL Research University, CNRS, Universit\'e Paris Diderot, Sorbonne Paris Cit\'e, 5 Place Jules Janssen, 92195 Meudon, France}
\author{M. Sereno}
\affiliation{INAF, Osservatorio Astronomico di Bologna, via Piero Gobetti 93/3, I-40129 Bologna, Italy}
\affiliation{Dipartimento di Fisica e Astronomia, Universit\`a di Bologna, via Piero Gobetti 93/2, I-40129 Bologna, Italy}
\author{S. Amodeo}
\affiliation{LERMA, Observatoire de Paris, PSL Research University, CNRS, Sorbonne Universit\'e, UPMC Univ. Paris 6, F-75014 Paris, France}
\author{M.-A. Breton}
\affiliation{LUTH, Observatoire de Paris, PSL Research University, CNRS, Universit\'e Paris Diderot, Sorbonne Paris Cit\'e, 5 Place Jules Janssen, 92195 Meudon, France}
\author{V. Ghirardini}
\affiliation{INAF, Osservatorio Astronomico di Bologna, via Piero Gobetti 93/3, I-40129 Bologna, Italy}
\affiliation{Dipartimento di Fisica e Astronomia, Universit\`a di Bologna, via Piero Gobetti 93/2, I-40129 Bologna, Italy}
\author{D. Eckert}
\affiliation{Max-Planck-Institute for Extraterrestrial Physics (MPE), Giessenbachstrasse 1, 85748 Garching, Germany}

\begin{abstract}
We present a new cosmological probe for galaxy clusters, the halo sparsity. This characterises halos in terms of the ratio of halo masses measured at two different radii and carries cosmological information encoded in the halo mass profile. Building upon the work of \citet{Balmes2014} we test the properties of the sparsity using halo catalogs from a numerical N-body simulation of ($2.6$ Gpc/h)$^3$ volume with $4096^3$ particles. We show that at a given redshift the average sparsity can be predicted from prior knowledge of the halo mass function. This provides a quantitative framework to infer cosmological parameter constraints using measurements of the sparsity of galaxy clusters. We show this point by performing a likelihood analysis of synthetic datasets with no systematics, from which we recover the input fiducial cosmology. We also perform a preliminary analysis of potential systematic errors and provide an estimate of the impact of baryonic effects on sparsity measurements. We evaluate the sparsity for a sample of 104 clusters with hydrostatic masses from X-ray observations and derive constraints on the cosmic matter density $\Omega_m$ and the normalisation amplitude of density fluctuations at the $8$ Mpc h$^{-1}$ scale, $\sigma_8$. Assuming no systematics, we find $\Omega_m=0.42\pm 0.17$ and $\sigma_8=0.80\pm 0.31$ at $1\sigma$, corresponding to $S_8\equiv \sigma_8\sqrt{\Omega_m}=0.48\pm 0.11$. Future cluster surveys may provide opportunities for precise measurements of the sparsity. A sample of a few hundreds clusters with mass estimate errors at a few percent level can provide competitive cosmological parameter constraints complementary to those inferred from other cosmic probes. 
\end{abstract}

\keywords{X-rays: galaxies: clusters --- cosmology: theory --- cosmology: cosmological parameters --- methods: numerical}

\section{Introduction}\label{intro}
In the standard bottom-up scenario of cosmic structure formation, initially small dark matter (DM) density fluctuations grow under gravitational instability to eventually form at later times virialized stable objects, the halos. It is inside these gravitationally bounded clumps of DM that baryonic gas falls in to form the visible structures we observe in the universe. Today, the most massive halos host large clusters of galaxies resulting from the hierarchical merging process of smaller mass halos formed at earlier times. Since their assembly depends on the matter content of the Universe, the state of cosmic expansion and the initial distribution of matter density fluctuations, there is a consensus that observations of galaxy clusters can provide a wealth of cosmological information \citep[see e.g.][for a review of galaxy cluster cosmology]{Allen2011,KravtsovBorgani2012}. 

Galaxy clusters can be observed through a variety of probes, such as the detection of the X-ray emission of the intra-cluster gas \citep[e.g.][]{Vikhlinin2005,MACS,MCXC,XXL}, the Sunyaev-Zeldovich effect in the microwave \citep[e.g.][]{Staniszewski2009,ACT,SPT,PLANCK}, the distribution of the member galaxies in the optical and IR bands \citep{MAXBCG,Rykoff2014}, and the distorsion of the background galaxies induced by the halo gravitational potential \citep[e.g.][]{Umetsu2011,Postman2012,Hoekstra2012}.

Due to the highly non-linear nature of the gravitational collapse driving the formation of DM halos, theoretical model predictions, which are necessary to interpret the data and infer cosmological parameter constraints have been mainly obtained through cosmological simulations. A remarkable result of these studies is the fact that DM halos exhibit a universal density profile well approximated by the Navarro-Frenk-White formula \citep{NFW}. This entirely characterises the halo profile in terms of the halo mass ${\rm M}$ and the {\it concentration} parameter $c$. Numerical simulations have shown that the concentration encodes cosmological information. In particular, it has been found that the median concentration of an ensemble of halos is a power-law function of the halo mass with the overall amplitude of the relation varying with redshift and cosmology \citep[see e.g.][]{Bullock2001,Zhao2003,Dolag2004,Zhao2009,Giocoli2011}. This has suggested that estimates of the concentration and halo mass from a sample of galaxy clusters can provide constraints on cosmological models \citep[see e.g][for a cosmological data analysis using cluster concentration-mass measurements]{Ettori2010}. 

However, several factors can limit the use of the cluster concentration as cosmological proxy. On the one hand astrophysical effects may alter the original $c-{\rm M}$ relation and introduce a systematic bias in the cosmological analysis \citep[see e.g.][]{Duffy2010,Mead2010,King2011}. On the other hand, theoretical model predictions, despite recent progress \citep[see e.g.][]{Diemer2015,Correa2015,Klypin2016,Ludlow2016,Renneby2017}, have yet to converge into a single model capable of reproducing the ensemble of numerical results currently available for different cosmological scenarios \citep{MeneghettiRasia2013}.

Another limiting factor may result of the large intrinsic dispersion of the halo concentration. N-body simulation studies have found a significant scatter of the concentration at fixed halo mass \citep{Bullock2001,Wechsler2002}. For example \citet{Maccio2006} have found $\sigma_{\ln c}\approx 0.25$, while \citet{Bhattacharya2013} quotes a scatter $\sigma_{\ln c}\approx 0.33$. A similar result has been found in \citet{Diemer2015}, which quotes $\sigma_{\ln c}\approx 0.37$, while a smaller value was only found for a sample of relaxed halos. Accounting for such a large intrinsic dispersion may strongly relax cosmological parameter constraints from measurements of the concentration-mass relation. 

A further point of concern is the case of very massive clusters. These are often easier to detect because they are very luminous and rich. Nonetheless, because of their recent formation they are also more likely to be perturbed by the presence of other structures that are still in the process of merging with the main DM halo. In such a case, the halo density profile may deviate from the NFW formula and the concentration parameter no longer encodes information of the halo mass distribution and its cosmological dependence. 

Finally, the measurement of the mass-concentration relation is strongly affected by selection effects as shown by \citet{Serenoetal2015}.

In \citet{Balmes2014}, two of the authors have introduced the concept of halo {\it sparsity}, a directly measurable proxy of the DM halo mass profile that overcomes most of the limitations described above. In this work, we present a detailed study of the validity of the halo sparsity as a new cosmological probe. As a proof-of-concept application, we specifically focus on sparsity measurements based on hydrostatic mass estimates from X-ray cluster observations. We show that the redshift evolution of the average halo sparsity carries cosmological information which can be retrieved from prior knowledge of the halo mass function. To this purpose we perform a likelihood analysis over a set of ideal sparsity data with no systematic errors from which we recover the input fiducial cosmology. We discuss various sources of systematic uncertainty. Using results from state of the art simulations of galaxy clusters we show that mass bias effects due to baryonic feedback processes alter the sparsity of massive systems at a few percent level. When analysing cluster sparsity, this source of systematic error is therefore subdominant with respect to that affecting mass estimates from currently available cluster datasets. As a first cosmological application, we perform a cosmological parameter inference analysis of sparsity measurements based on hydrostatic mass estimates of a sample of X-ray clusters from {\it XMM} and {\it Chandra} observations. 

The article is organised as follows. In Section~\ref{sec_sparsity}, we review the basic properties of the halo sparsity and test its validity as a cosmological proxy. In Section~\ref{sys}, we discuss several sources of systematic errors that can affect the sparsity data analysis. In Section~\ref{data}, we present the cosmological parameter constraints inferred from sparsity measurements of a sample of X-ray clusters. In Section~\ref{forecast}, we perform a cosmological parameter forecast for sparsity data expected from future X-ray cluster surveys. Finally, in Section~\ref{conclu} we present our conclusions.

\section{Dark Matter Halo Sparsity}\label{sec_sparsity}

\subsection{Definition \& Properties}
The sparsity of a halo is defined as the ratio of the halo mass enclosing two different overdensities $\Delta_1$ and $\Delta_2$ \citep{Balmes2014}:
\begin{equation}\label{sparsity}
s_{\Delta_1,\Delta_2}\equiv \frac{{\rm M}_{\Delta_1}}{{\rm M}_{\Delta_2}},  
\end{equation}
where $\Delta_1<\Delta_2$ and ${\rm M}_{\Delta}$ is the mass enclosed in a sphere of radius $r_{\Delta}$ containing an overdensity $\Delta$ with respect to the critical density $\rho_c$ or the mean background density $\rho_m$. In the following we will consider $\rho_c$, however as shown in \citet{Balmes2014} the general properties of the sparsity are independent of such a choice. Notice that from Eq.~(\ref{sparsity}) we can also interpret the sparsity of a halo as a measure of the mass excess between $r_{\Delta_1}$ and $r_{\Delta_2}$ ($\Delta{\rm M}={\rm M}_{\Delta_1}-{\rm M}_{\Delta_2}$) relative to the mass enclosed in the inner radius $r_{\Delta_2}$, i.e. $s_{\Delta_1,\Delta_2}=\Delta{\rm M}/{\rm M}_{\Delta_2}+1$. 

It is easy to show that there is a one-to-one correspondence between the halo sparsity and the concentration parameter (assuming that the halo follows the NFW profile). For instance to be conform with the standard definition of concentration let us set $\Delta_1=200$ and let be $\Delta_2=\Delta$, using the NFW formula we can write the sparsity as:
\begin{equation}\label{conspar}
s_{\Delta}^{-1}\equiv x_{\Delta}^3\frac{\Delta}{200}=\frac{\ln{(1+c_{200}\,x_{\Delta})}-\frac{c_{200}\,x_{\Delta}}{1+c_{200}\,x_{\Delta}}}{\ln{(1+c_{200})}-\frac{c_{200}}{1+c_{200}}},
\end{equation}
where $x_{\Delta}=r_{\Delta}/r_{200}$ and $c_{200}=r_{200}/r_s$ with $r_s$ the scale radius of the NFW profile. For a given set of values of the concentration $c$ and overdensity $\Delta$, the above equation can be solved numerically to obtain $x_{\Delta}$ and thus the corresponding value of $s_{\Delta}$. However, notice that in defining the sparsity as in Eq.~(\ref{sparsity}) no explicit assumption has been made concerning the form of the halo density profile. \citet{Balmes2014} have shown that this is sufficient to characterise the mass profiles of halos even when their density profile deviates from NFW. 

A key feature of the halo sparsity is the fact that its ensemble average value at a given redshift is nearly independent of the halo mass ${\rm M}_{\Delta_1}$ (even if some of the halos in the ensemble have profiles which deviates from NFW), but depends on the underlying cosmological model with a scatter that is much smaller than that of the halo concentration. Because of this it can provide a robust cosmological proxy, without requiring any modelling of the halo density profile. 

Another important characteristic of the halo sparsity is that its independence on ${\rm M}_{\Delta_1}$ implies that the ensemble average value can be predicted from prior knowledge of the halo mass function at two different mass overdensities. In fact, let us consider the equality
\begin{equation}
\frac{dn}{d{\rm M}_{\Delta_2}}=\frac{dn}{d{\rm M}_{\Delta_1}}\frac{d{\rm M}_{\Delta_1}}{d{\rm M}_{\Delta_2}}=\frac{dn}{d{\rm M}_{\Delta_1}}s_{\Delta_1,\Delta_2}\frac{d\ln{{\rm M}_{\Delta_1}}}{d\ln{{\rm M}_{\Delta_2}}},
\end{equation}
where $dn/d{\rm M}_{\Delta_1}$ and $dn/d{\rm M}_{\Delta_2}$ are the mass functions at $\Delta_1$ and $\Delta_2$ respectively. We can rearrange the above relation and integrate over the halo ensemble mass range to derive the relation between the average inverse halo masses at two different overdensities. Since the sparsity is independent of the halo mass, it can be taken out of the integration on the right-hand-side such that
\begin{equation}
\int_{{\rm M}^{\rm min}_{\Delta_2}}^{{\rm M}^{\rm max}_{\Delta_2}}\frac{dn}{d{\rm M}_{\Delta_2}}d\ln{{\rm M}_{\Delta_2}}=\langle s_{\Delta_1,\Delta_2}\rangle\int_{\langle s_{\Delta_1,\Delta_2}\rangle {\rm M}^{\rm min}_{\Delta_2}}^{\langle s_{\Delta_1,\Delta_2}\rangle {\rm M}^{\rm max}_{\Delta_2}}  \frac{dn}{d{\rm M}_{\Delta_1}}d\ln{{\rm M}_{\Delta_1}},\label{sparpred}
\end{equation}
this equation can be solved numerically for $\langle s_{\Delta_1,\Delta_2}\rangle$ given prior knowledge of $dn/d{\rm M}_{\Delta_1}$ and $dn/d{\rm M}_{\Delta_2}$ respectively. As shown in \citet{Balmes2014}, this reproduces with great accuracy the mean sparsity inferred from N-body halo catalogs. Indeed, this is a direct advantage over predicting the median concentration, since the cosmological and redshift dependence of the mass function are easier to model than the concentration, as they involve a reduced set of assumptions. Moreover, since the sparsity is a mass ratio, it is reasonable to expect that it will be less affected by a constant systematic bias which may affect cluster mass measurements. Also, notice that selection effects can be included in Eq.~(\ref{sparpred}) by convolving the integrands with the appropriate selection function. We will discuss both these points in detail in Section~\ref{sys}.

A last remark concerns the choice of $\Delta_1$ and $\Delta_2$ provided that $\Delta_1<\Delta_2$. As noticed in \citet{Balmes2014} the larger the difference, the greater the amplitude of the cosmological signal. However, the values of $\Delta_1$ and $\Delta_2$ cannot be chosen to be arbitrarily different since the properties of the sparsity discussed above remain valid only in a limited interval which can be determined by physical considerations. A lower bound on $\Delta_1$ can be inferred by the fact that for very small overdensities the identification of a halo as an individual object can be ambiguous, thus suggesting $\Delta_1\gtrsim 100$. On the other hand, the range of values for $\Delta_2$ can be deduced by the fact that at very large overdensities baryonic processes may alter the DM distribution within the inner core of halos. These are largely subdominant if one conservatively assumes $\Delta_2$ in the range $\Delta_1<\Delta_2\lesssim 2000$. Within this interval of values one can set $\Delta_1$ and $\Delta_2$ depending on the availability of optimal mass measurements.

\subsection{N-body Simulation Sparsity Tests}\label{nbodytest}
In \citet{Balmes2014} the properties of the halo sparsity have been tested using halo catalogs from the Dark Energy Universe Simulations\footnote{http://www.deus-consortium.org/deus-data/} (DEUS) with masses defined with respect to the background density covering the mass range $10^{12}<{\rm M}_{200m}\,[h^{-1}\,{\rm M}_{\odot}]<10^{15}$. 

Here, we perform an analysis using halos identified with the Spherical Overdensity (SOD) halo detection algorithm \citep{LaceyCole1994} in one of the simulations of the RayGalGroupSims suite \citep{Raserainprep} with masses defined with respect to the critical density. Since we are interested in the application to X-ray clusters we specifically focus on masses at overdensity $\Delta_1=500c$ and $\Delta_2=1000c$ from which we derive estimates of the halo sparsity $s_{500,1000}$. For completeness, we also consider halo masses measured at $\Delta_1=200c$ and show that the properties of the halo sparsity also hold for $s_{200,500}$ and $s_{200,1000}$ respectively.  

The cosmological model of the RayGalGroupSims simulation considered here is a flat $\Lambda$CDM with parameters 
set consistently with WMAP-7 year data analysis \citep{Komatsu2011}: $\Omega_m=0.2573$, $\Omega_b=0.04356$, $h=0.72$, $n_s=0.963$ and $\sigma_8=0.801$. The simulation consists of a ($2.625$ Gpc $h^{-1}$)$^3$ volume with $4096^3$ particles corresponding to particle mass resolution $m_p=1.88\cdot 10^{10}$ M$_\odot$ h$^{-1}$. 

Halos are first detected using the SOD algorithm with overdensity set to $\Delta=200c$ and centred on the location of maximum density. For each halo we computed SOD masses at $\Delta=200c,500c$ and $1000c$ respectively and estimated the corresponding halo sparsities. In order to avoid mass resolution effects we have taken a conservative mass cut and considered only halos with more than $10^4$ particles.

\begin{figure}
\centering
\includegraphics[width=.45\textwidth]{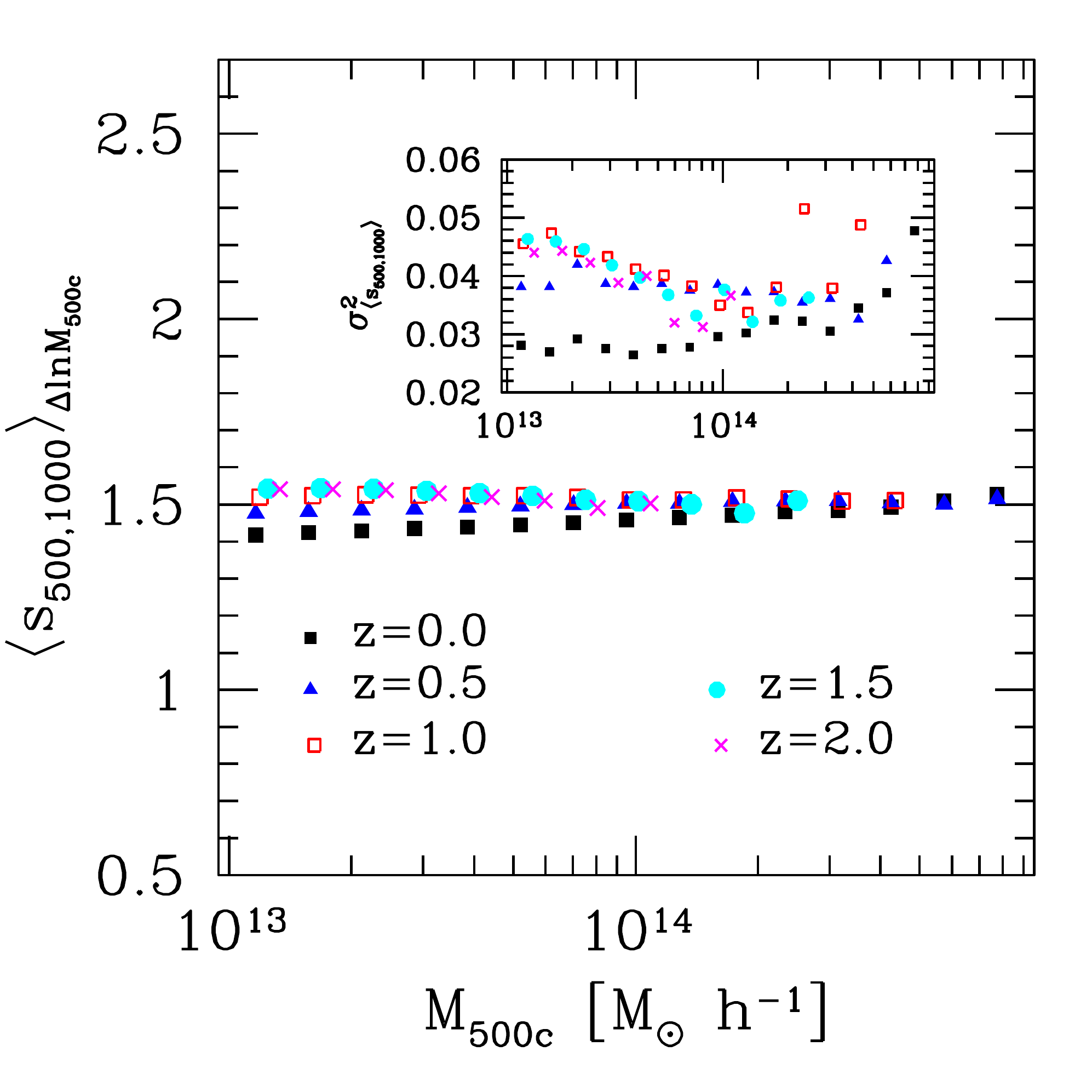}
\caption{Average halo sparsity as function of ${\rm M}_{500c}$ from SOD halo catalogs at $z=0$ (black filled squares), $z=0.5$ (blue filled triangles), $1.0$ (red empty squares), $1.5$ (cyan solid circles) and $2.0$ (magenta crosses) in mass bins of size $\Delta\ln{{\rm M}_{500c}}=0.3$. The inset plot shows the variance of the halo sparsity in the same mass bins as function of ${\rm M}_{500c}$ for the different redshifts.}
\label{fig:avs_m500c}
\end{figure}

In Fig.~\ref{fig:avs_m500c} we plot the average halo sparsity $\langle s_{500,1000}\rangle_{\Delta \ln{{\rm M}_{500c}}}$ in mass bins of size $\Delta \ln{{\rm M}_{500c}}=0.3$ (containing more than $20$ halos) as function of ${\rm M}_{500c}$ at $z=0,0.5,1.0,1.5$ and $2.0$, while in the inset plot we show the associated variance. As we can see, $\langle s_{500,1000}\rangle_{\Delta \ln{{\rm M}_{500c}}}$ remains constant to very good approximation across the full mass and redshift range. As far as the scatter is concerned we find the standard deviation to be $\lesssim 20\%$ level, consistently with the findings of \citet{Balmes2014}.

\begin{figure}[t]
\centering
\includegraphics[width=.45\textwidth]{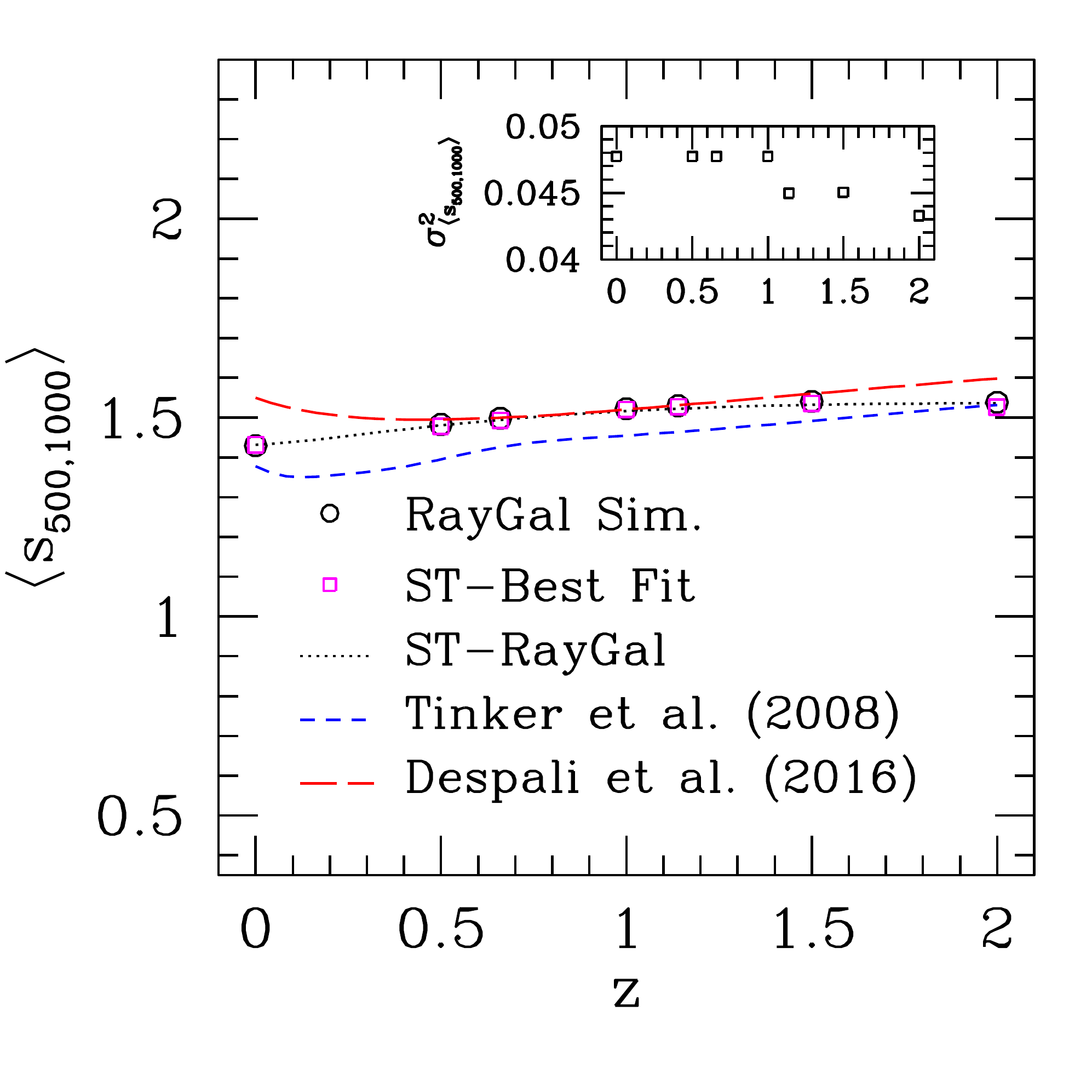}
\caption{Average sparsity as function of redshift for halos with ${\rm M}_{500c}> 10^{13}$ M$_{\odot}$ h$^{-1}$. The black empty circles correspond to the average sparsity measured from the halo catalogs at the redshift snapshots of the RayGalGroupSims run. The magenta empty squares correspond to the average sparsity prediction obtained by solving Eq.~(\ref{sparpred}) assuming a ST mass function with coefficients best-fitting the halo mass function of the RayGalGroupSims halo catalogs at ${\rm M}_{500c}$ and ${\rm M}_{1000c}$ respectively. The black dotted line corresponds to the prediction from Eq.~(\ref{sparpred}) using the ST-RayGal mass function, while the blue short-dashed line and the red long-dashed line correspond to the predictions obtained by assuming the mass function from \citet{Tinker2008} and \citet{Despali2016}, respectively. The inset plot shows the variance of the halo sparsity from the SOD halo catalogs with ${\rm M}_{500c}\gtrsim 10^{13}$ M$_{\odot}$ h$^{-1}$ at the different redshift snapshots.}
\label{fig:sparsz_th_vs_nh}
\end{figure}

Let us now test the validity of Eq.~(\ref{sparpred}) in predicting the redshift evolution of the ensemble average sparsity. In Fig.~\ref{fig:sparsz_th_vs_nh}, the solid black circles are the average sparsity values  obtained from the RayGalGroupSims halo catalogs at $z=0,0.5,0.66,1.0,1.14,1.5$ and $2.0$ respectively. These have been computed for each halo catalog by averaging the sparsity of halos with ${\rm M}_{500c}\gtrsim 10^{13}$ M$_{\odot}$ h$^{-1}$. The solid magenta squares are the average sparsity values at the redshifts of the halo catalogs obtained by solving Eq.~(\ref{sparpred}) where we have assumed the Sheth-Tormen (ST) formula \citep{Sheth1999} with coefficients best-fitting the numerical halo mass functions at $\Delta_1=500c$ and $\Delta_2=1000c$ respectively (see Appendix~\ref{app_hmf} for a detailed description of the mass function calibration). As we can see in Fig.~\ref{fig:sparsz_th_vs_nh}, the predictions from Eq.~(\ref{sparpred}) overlap with the average sparsity values directly estimated from the halos in the simulation catalogs with relative differences $<0.1\%$ level\footnote{In solving Eq.~(\ref{sparpred}) we have set ${\rm M}^{\rm min}_{1000c}=2\cdot10^{13}$ M$_{\odot}$ h$^{-1}$ consistently with the mass limit of our halo catalogs, while the upper limit of the integration interval can be set to an arbitrarily large number. Nevertheless, as the average sparsity remains approximately constant with mass, we have verified that the solution of Eq.~(\ref{sparpred}) is largely independent of the specific choice of ${\rm M}^{\rm min}_{1000c}$.}.

In order to interpolate predictions of the sparsity at redshifts other than those tested by the simulation snapshots we have performed a quadratic fit of the ST best-fit coefficients as function of $x\equiv\log_{10}(\Delta_c/\Delta_{\rm vir}(z))$ for $\Delta_c=500$ and $1000$ respectively, see Eq.~(\ref{coefm500c})-(\ref{coefm1000c}) in Appendix~\ref{app_hmf}. As suggested by \citet{Despali2016}, parametrising the ST-coefficients in terms of $x$ aims to capture the redshift and cosmology dependence of the mass function, though from the work of \citet{Courtin2011} it is clear that this may not be sufficient to model dependencies beyond the $\Lambda$CDM scenario.  Hereafter, we will refer to the ST formula with coefficients given by Eq.~(\ref{coefm500c})-(\ref{coefm1000c}) as the ST-RayGal mass function, the corresponding average sparsity prediction from Eq.~(\ref{sparpred}) is shown in Fig.~\ref{fig:sparsz_th_vs_nh} as black dotted line. We find differences with respect to the N-body measurements to be at sub-percent level. 

In Fig.~\ref{fig:sparsz_th_vs_nh} we also plot the average sparsity prediction from Eq.~(\ref{sparpred}) obtained by assuming the mass function from \citet{Tinker2008} and \citet{Despali2016} respectively. In the former case we can see systematic deviations up to $\sim 10\%$ level with respect to the N-body estimates that decrease from low to high redshifts. In the latter case differences are within $1\%$ level in the range $0.5<z<1.5$, while they increase up to $\sim 10\%$ level at lower and higher redshifts. Such discrepancies are due to differences in the parameterisations of the halo mass function which have been calibrated to halo catalogs from simulations of different cosmological models, volumes and mass resolutions. 

Compared to the simulations used in \citet{Tinker2008,Despali2016}, the RayGalGroupSims simulation covers a larger volume with greater mass resolution. This provides a better calibration of the ST formulae. As it can be seen in Fig.~\ref{hmf_app} in Appendix~\ref{app_hmf}, we find logarithmic differences well within $5\%$ level. On the other hand, it is worth remarking that we have tested the validity of the ST-RayGal mass function to a set of cosmological simulations with parameters which are not too different from those of the $\Lambda$CDM best-fit model to the WMAP-7 year data (see discussion at the end of Appendix~\ref{app_hmf}). Consequently, we are not guaranteed that the ST-RayGal parameterisation can fully capture the cosmological parameter dependence of the halo mass function and hence that of the halo sparsity for parameter values which are far from the concordance $\Lambda$CDM model. Such uncertainty can indeed introduce systematic errors in the sparsity analysis, a point which we will discuss in detail in Section~\ref{sys}. Here, we are not in a position to solve this issue in a conclusive manner. Hence, we simply opt to quote the results obtained assuming the ST-RayGal parameterisation and that from \citet{Despali2016}. We will refer to the latter case as ST-Despali. 

\begin{figure*}
\centering
\includegraphics[width=.4\textwidth]{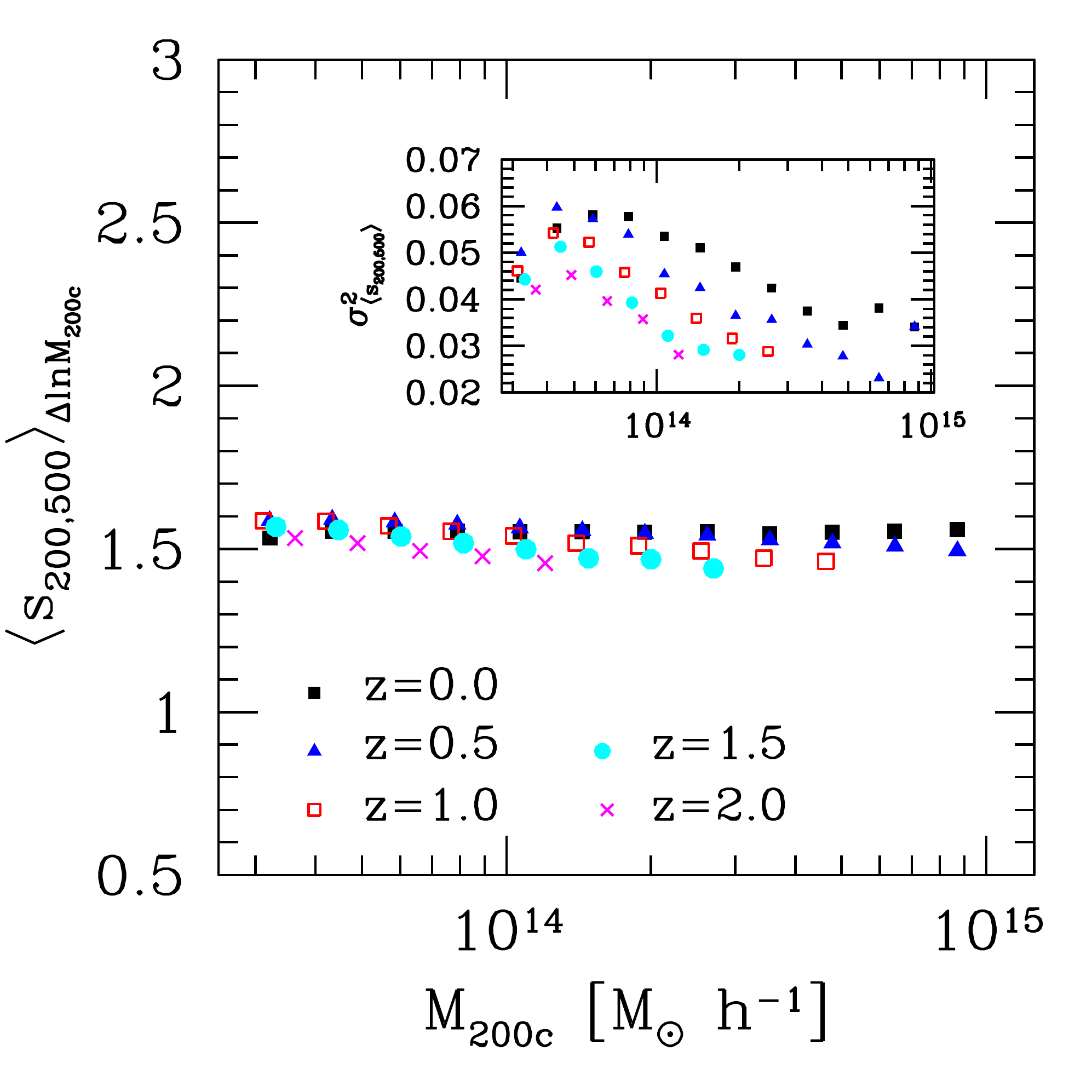}
\includegraphics[width=.4\textwidth]{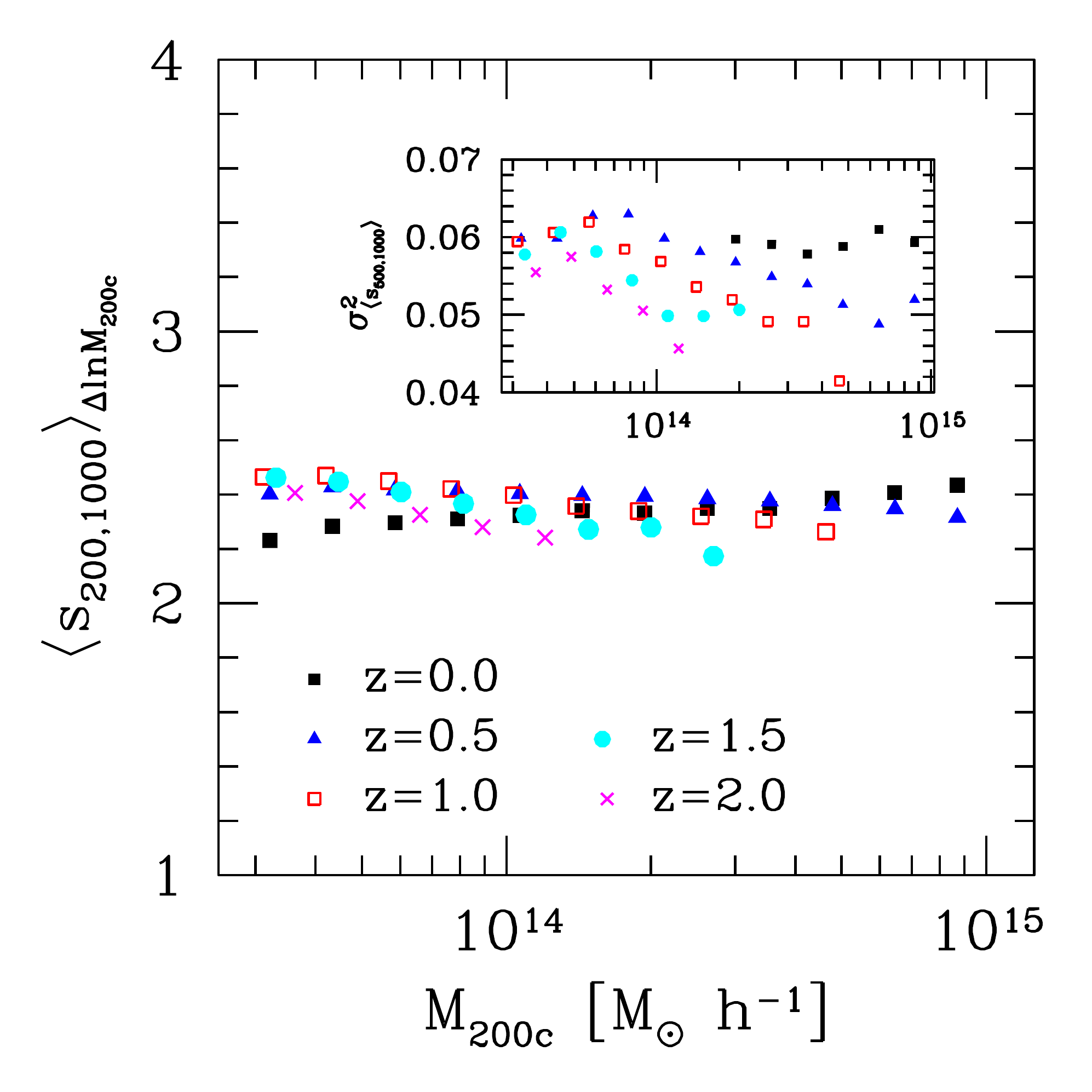}
\includegraphics[width=.4\textwidth]{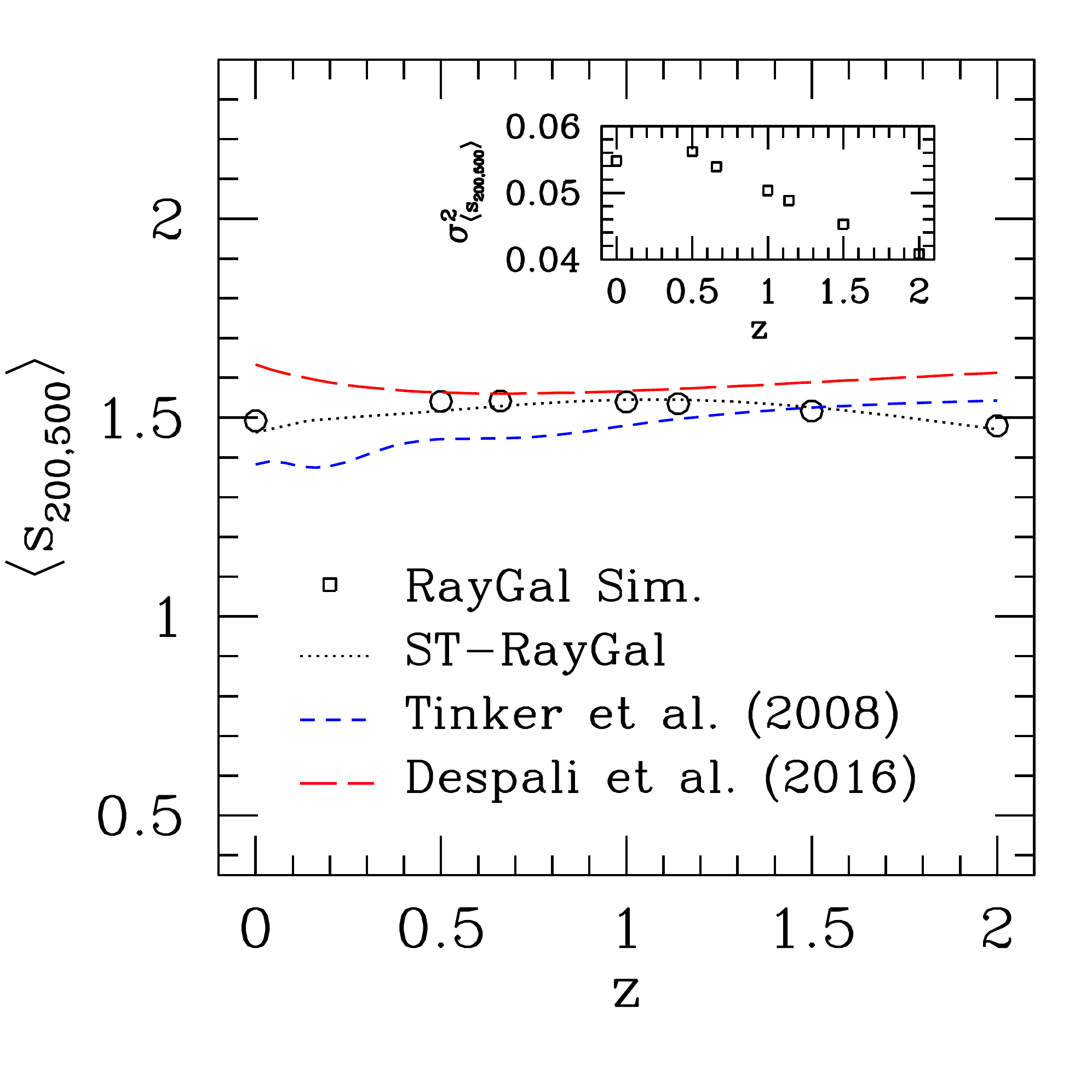}
\includegraphics[width=.4\textwidth]{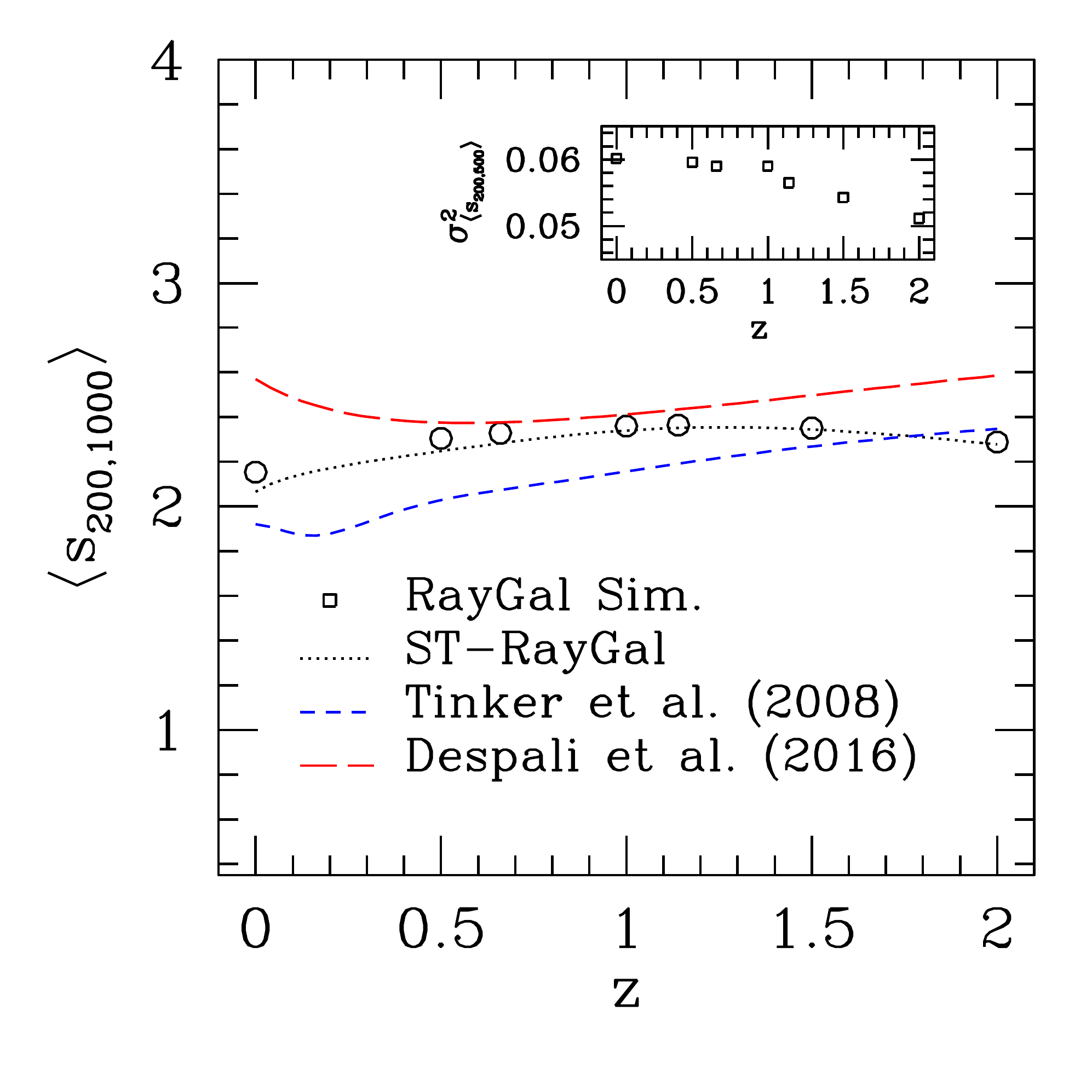}
\caption{Top panels: average halo sparsity $\langle s_{200,500}\rangle$ (left panel) and $\langle s_{200,1000}\rangle$ as function of ${\rm M}_{200c}$ at $z=0,0.5,1,1.5$ and $2.0$ (legend as in Fig.~\ref{fig:avs_m500c}) in mass bins of size $\Delta\ln{{\rm M}_{200c}}=0.3$. The inset plot shows the variance of the halo sparsity in the same mass bins as function of ${\rm M}_{200c}$ for the different redshifts. Bottom panels: average halo sparsity $\langle s_{200,500}\rangle$ (left panel) and $\langle s_{200,1000}\rangle$ as function of redshift for halos (legend as in Fig.~\ref{fig:sparsz_th_vs_nh}) with ${\rm M}_{500c}> 10^{13}$ M$_{\odot}$ h$^{-1}$ consistently with the mass cut adopted for the $s_{500,1000}$ case. The inset plot shows the variance of the halo sparsity at the different redshift snapshots.}
\label{sparsity_other_deltas}
\end{figure*}

The properties of the halo sparsity summarised by the trends shown in Fig.~\ref{fig:avs_m500c} and Fig.~\ref{fig:sparsz_th_vs_nh} also hold for other sparsity definitions. This can be seen in Fig.~\ref{sparsity_other_deltas}, where we plot $\langle s_{200,500}\rangle$ and $\langle s_{200,1000}\rangle$ as function of $M_{200c}$ and redshift respectively. These plots suggest that sparsity estimations from mass measurements at $\Delta=200c$, such as those provided by gravitational lensing observations that probe clusters at larger radii than X-ray measurements, can also provide a viable proxy of the mass distribution in clusters.

\begin{figure}[t]
\centering
\includegraphics[width=.45\textwidth]{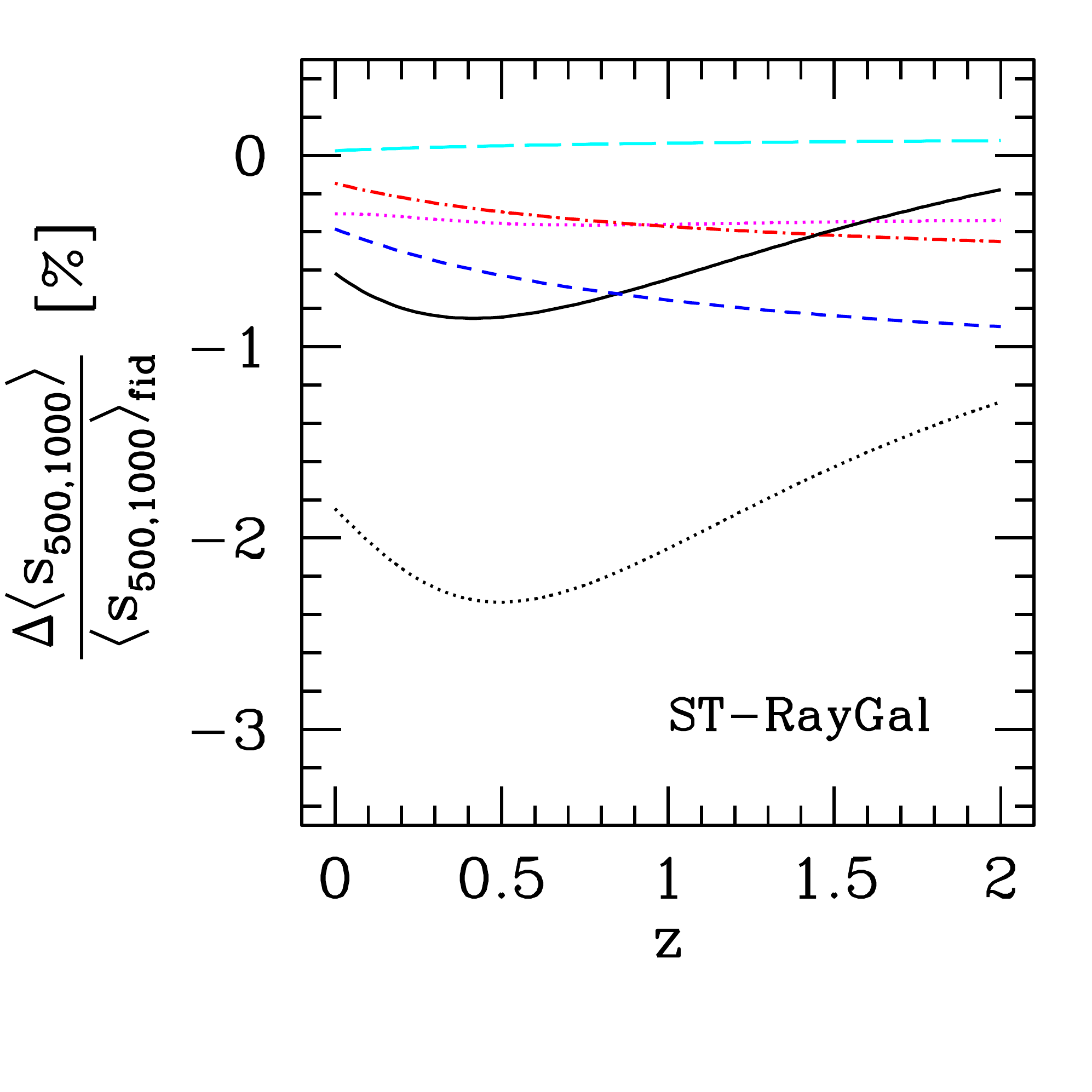}
\includegraphics[width=.45\textwidth]{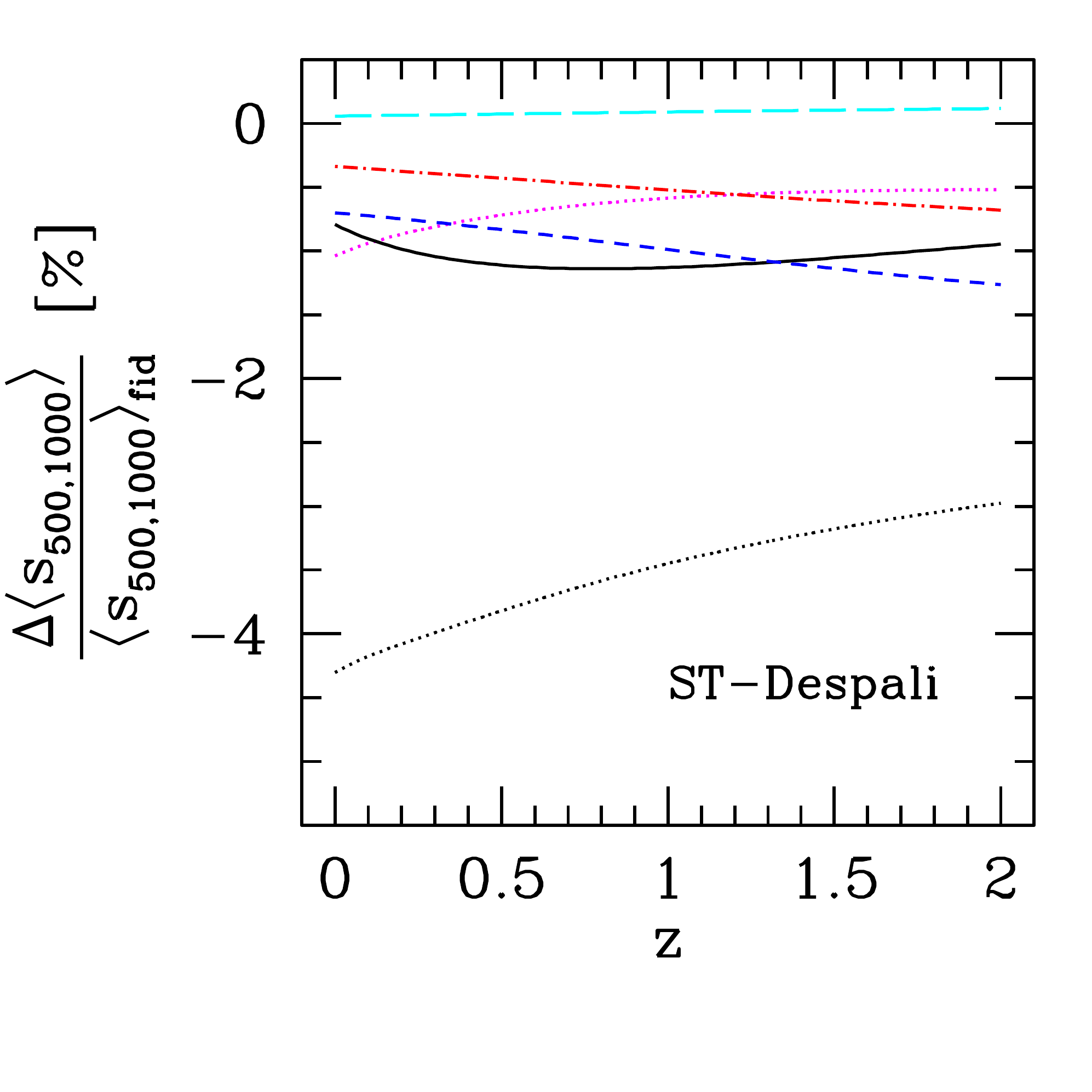}
\caption{\label{fig:cosmo_derivs} Relative variation of the average halo sparsity as function of redshift for a $5\%$ variation of the cosmological parameters around the fiducial {\it Planck} values. The various lines represent variations with respect to $\sigma_8$ (black solid line), $n_s$ (blue short-dashed line), $\Omega_m$ (magenta dotted line), $h$ (red dotted short-dashed line), $\Omega_b\,h^2$ (cyan long-dashed line) and $\sigma_8\sqrt{\Omega_m}$ (black dotted line). In the upper panel we plot the relative variation of the average sparsity obtained assuming the ST-RayGal mass function in Eq.~(\ref{sparpred}), while that assuming the ST-Despali mass function is shown in the lower panel.}
\end{figure}

\subsection{Cosmological Parameter Dependence}\label{cosmodep}
The dependence of the halo sparsity on the underlying cosmological model has been studied in \citet{Balmes2014} using N-body halo catalogs from DEUS project simulations \citep{Alimi2010,Rasera2010,Courtin2011}. \citet{Balmes2014} have shown that the average value of the sparsity at a given redshift correlates with the linear growth factor of the underlying cosmological model. This can be qualitatively understood in terms of the relation between the growth of structures and the mass assembly of halos. In particular, at any given time, models which form structures earlier will assemble on average more halo mass at large overdensities (inner radii) than those which form structures at later times, thus resulting in smaller values of the average sparsity. In terms of the cosmological model parameters, this implies for instance that the larger the cosmic matter density $\Omega_m$ or the amplitude of the fluctuations on the $8$ Mpc h$^{-1}$ scale $\sigma_8$ and the smaller the average sparsity value. 

Here, we do not intend to repeat the analysis of \citet{Balmes2014}, instead we use Eq.~(\ref{sparpred}) to evaluate the relative change of the average sparsity with respect to a fiducial cosmological model for a positive variation of the cosmological parameters.
 
We assume as fiducial cosmology a flat $\Lambda$CDM model with parameters set to the best-fit values from the {\it Planck} cosmological data analysis of Cosmic Microwave Background (CMB) anisotropy spectra (TT,TE,EE+lowP) \citep{PlanckCosmo}: $\Omega_m=0.3156$, $\Omega_b\,h^2=0.02225$, $h=0.6727$, $\sigma_8=0.831$, $n_s=0.9645$. 

We compute $\langle s_{500,1000}\rangle$ from Eq.~(\ref{sparpred}) assuming the ST-RayGal and ST-Despali mass functions respectively\footnote{In computing the mass function we evaluate the linear matter power spectrum of the underlying cosmological model using the approximated formulae from \citet{Hu1998}. We have verified that using power spectra from numerical solutions of linear perturbation equations, such as those given by the CAMB code \citep{CAMB}, leads to sub-percent difference in the predicted value of the average sparsity.}.

In Fig.~\ref{fig:cosmo_derivs} we plot $\Delta{\langle s_{500,1000}\rangle}/{\langle s_{500,1000}\rangle}_{\rm fid}$ as function of redshift in the case of the ST-RayGal mass function (upper panel) and ST-Despali mass function (lower panel). Independently of the adopted mass function parametrisation, we can see that the variation of the average sparsity is negative for a positive variation of the cosmological parameters, except $\Omega_b\,h^2$. This is essentially because increasing the value of $\sigma_8$, $\Omega_m$, $n_s$ and $h$ causes structures to form at earlier times, and consequently assemble more halo mass at larger overdensities, which results into smaller values of the average sparsity. This is not the case for a positive variations of $\Omega_b\,h^2$. In fact, as we have assumed a flat geometry, increasing the value of $\Omega_b\,h^2$ corresponds to decreasing the value of $\Omega_m$ at $h$ constant. In such a case structures form later than in models with smaller values of $\Omega_b\,h^2$, halos assemble on average less mass at larger overdensities, thus leading to larger values of the average sparsity.

The trends shown in Fig.~\ref{fig:cosmo_derivs} provide an estimate of the sensitivity of the average sparsity to the different cosmological parameters. In the ST-RayGal case we can see that a change in the value of $\sigma_8$ produces the largest variation of the average sparsity in the redshift range $0<z\lesssim 1$. At higher redshifts a change in the value of $n_s$ causes the largest variation, while $\Omega_m$, $h$ and $\Omega_b h^2$ have smaller effects. A similar trend occurs in the ST-Despali case, though with different amplitudes for the different parameters. Overall, we can see that measurements of the average sparsity are most sensitive to $S_8=\sigma_8\sqrt{\Omega_m}$, consequently we can expect constraints on $\Omega_m$ and $\sigma_8$ to be degenerate along curves of constant $S_8$ values. 

It is worth noticing that the variations of the average sparsity predicted by the ST-Despali mass function are slightly larger in amplitude than that from the ST-RayGal. This suggests that cosmological constraints inferred by a sparsity analysis based on the ST-Despali mass function will provide systematically tighter bounds than those inferred assuming the ST-RayGal parameterisation. As already mentioned at the end of Section~\ref{nbodytest}, the uncertainties in the modelling of the halo mass function may induce a systematic error in the cosmological analysis of sparsity measurements. We will discuss this in detail in Section~\ref{sys}.

\subsection{Synthetic Data Analysis}\label{synthetic}
We now check the validity of the average sparsity as cosmological proxy. To this purpose we generate a set of synthetic average sparsity data and perform a cosmological parameter likelihood analysis to test whether we retrieve the input parameter values of the fiducial cosmology. As proof-of-concept, {here we limit ourselves} to ideal sparsity measurements and neglect any source of systematic uncertainty. Our goal is to show that the sparsity provides a viable cosmological observable.

We assume as fiducial model a flat $\Lambda$CDM scenario with parameters set to the {\it Planck} best-fit values quoted in Section~\ref{cosmodep}. We generate a sample of $N=15$ independent sparsity measurements in redshift bins of size $\Delta{z}=0.1$ over the range $0\le z \le 1.5$ by solving Eq.~(\ref{sparpred}) for a given mass function. We consider two separate configurations, one consisting of sparsity measurements with $1\%$ statistical errors and another with $20\%$ errors. This allows us to test whether degrading the statistical uncertainties has an impact in retrieving the fiducial model. We focus the parameter inference on $\sigma_8$ and $\Omega_m$, while assuming hard priors on the remaining cosmological parameters. We realise two independent analyses for the ST-RayGal and the ST-Despali mass functions respectively.

\begin{figure}[t]
\centering
\includegraphics[width=0.45\textwidth]{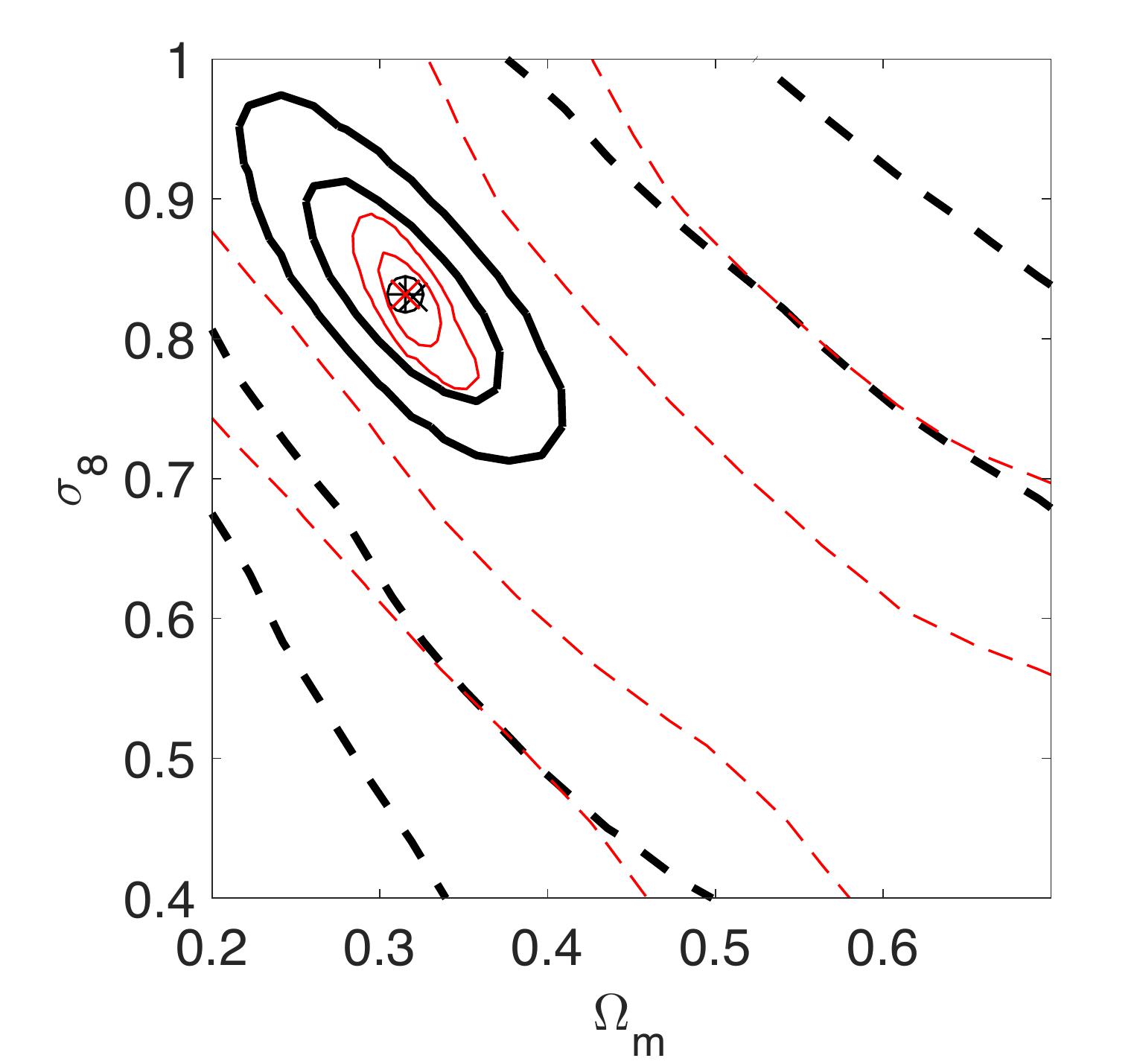}
\caption{$1$ and $2\sigma$ credibility contours in the $\Omega_m-\sigma_8$ plane obtained in the case of the ST-RayGal mass function (black thick lines) and the ST-Despali mass function (red thin lines). The solid lines correspond to constraints inferred assuming $1\%$ average sparsity errors, while the dashed contours correspond to the case with $20\%$ errors. The black crossed-circle indicates the fiducial model parameters, while the star markers indicate the best-fit values.}
\label{fig:like_test}
\end{figure}

We perform a Markov Chain Monte Carlo sampling of the likelihood function and evaluate the $\chi^2$:
\begin{equation}
\chi^2(\sigma_8,\Omega_m)=\sum_{i=1}^N\left[\frac{\langle s^i_{500,1000}\rangle-\langle s^{\rm th}_{500,1000}(z_i\vert\sigma_8,\Omega_m)\rangle}{\sigma_{\langle s_{500,1000}\rangle}}\right]^2,
\end{equation}
where $\langle s^i_{500,1000}\rangle$ is the $i$-th datapoint in the synthetic catalog at redshift $z_i$, $\sigma_{\langle s_{500,1000}\rangle}$ is the associated error and $\langle s^{\rm th}_{500,1000}\rangle$ is the theoretical model prediction given by Eq.~(\ref{sparpred}) assuming the same mass function parameterisation used to generate  the data.

The results are summarised in Fig.~{\ref{fig:like_test}} where we plot the $1$ and $2\sigma$ credibility contours in the plane $\Omega_m-\sigma_8$ which have been inferred assuming the ST-RayGal and ST-Despali mass function respectively. In both cases we find the best-fit model parameters to recover the {\it Planck} fiducial parameters at sub-percent level, independently of the assumed uncertainties on the synthetic dataset. On the other hand, we can see that the parameter constraints become much weaker in the case with $20\%$ statistical errors. As expected from the analysis presented in Section~\ref{cosmodep}, the analysis of the synthetic data performed using the ST-Despali mass function provides systematically tighter bounds on $\Omega_m-\sigma_8$ than the ST-RayGal case. 

Overall, this suggests that the average sparsity can be used as a cosmic probe. We will discuss extensively in the next Section the extent to which systematic errors can contaminate sparsity analyses.

\section{Systematic Errors}\label{sys}
In this section we present a preliminary evaluation of systematic errors potentially affecting cluster sparsity analyses. 

\subsection{Mass Function Parametrisation}
In Section~\ref{nbodytest} we have seen that key to predicting the halo sparsity is the correct modelling of the halo mass function. In particular, we have shown that Eq.~(\ref{sparpred}) recovers the average sparsity of the numerical halo catalogs from the RayGalGroupSims simulation provided that the parameterisation of the halo mass function for $M_{500c}$ and $M_{1000c}$ also reproduces the numerical halo abundances.

In order to asses the impact of the modelling of the mass function on the cosmological parameter inference from sparsity measurements we extend the synthetic data analysis presented in Section \ref{synthetic}. In particular, using the synthetic dataset generated by solving Eq.~(\ref{sparpred}) with the ST-RayGal mass function we perform a likelihood analysis assuming the ST-Despali mass function.

\begin{figure}[t]
\centering
\includegraphics[width=0.45\textwidth]{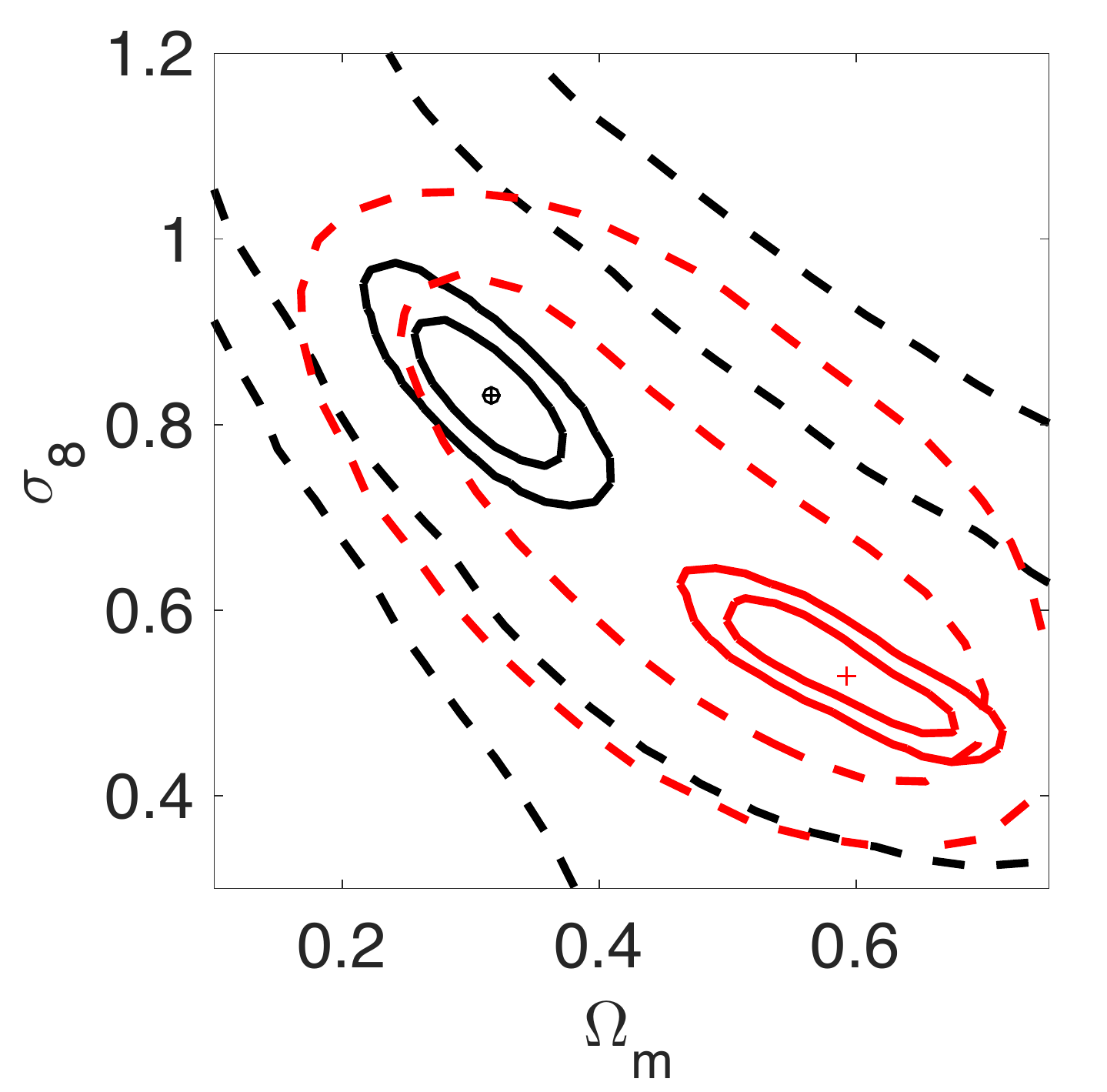}
\caption{\label{mfsys} $1$ and $2\sigma$ credibility contours in the $\Omega_m-\sigma_8$ plane from the likelihood analysis of average sparsity data with $1\%$ (solid lines) and $20\%$ (dashed lines) statistical errors generated by solving Eq.~(\ref{sparpred}) with the ST-RayGal mass function. The black lines correspond to the constraints shown in Fig.~\ref{fig:like_test} inferred assuming the ST-RayGal mass function. The constraints obtained assuming ST-Despali mass function are shown as red lines. The black circle indicates the fiducial model parameters, while the cross symbols indicate the best-fit parameter values for the different parameterisations.}
\end{figure}

In Fig.~\ref{mfsys} we plot the $1$ and $2\sigma$ credibility contours in $\Omega_m-\sigma_8$ plane assuming $1\%$ and $20\%$ statistical errors on the synthetic sparsity data respectively. For comparison we also plot the contours shown in Fig.~\ref{fig:like_test} obtained by assuming the ST-RaGal mass function. We can clearly see that assuming the ST-Despali mass function when the synthetic data have been generated with the ST-RayGal mass function results in a systematic off-set of the best-fit parameters. This bias is well above the statistical errors for sparsity measurements with $1\%$ statistical uncertainties. We can also notice that the contours differ according to the assumed mass function. This is not surprising given the fact that the mass function parameterisations have been calibrated to simulations of different volume and mass resolution and encode differently the dependence on the cosmological parameters.

As already mentioned the RayGalGroupSims simulation with a ($2.625$ Gpc $h^{-1}$)$^3$ volume and a mass resolution of $m_p=1.88\cdot 10^{10}$ M$_\odot$ h$^{-1}$ provides a better sampling of the high mass-end of the halo mass function than the simulation ensemble used for the calibration of the ST-Despali mass function. As an example, the largest volume simulation from the SBARBINE suite presented in \citep{Despali2016} consists of a ($2$ Gpc $h^{-1}$)$^3$ box with mass resolution $m_p=6.35\cdot 10^{11}$ M$_\odot$ h$^{-1}$. This impact the accuracy of the mass function calibration, a point that can be also inferred by comparing the amplitude of the logarithmic differences of the calibrated formulae to the numerical estimations. As shown in Fig.~\ref{hmf_app}, the ST-RayGal parametrisation reproduces the RayGalGroupSims mass function well within $5\%$ across the entire mass range probed by the simulation and in redshift interval $0\le z\le2$. In contrast the ST-Despali parametrisation shows differences with respect to the N-body results that at the high mass-end exceed the $5\%$ level in the same redshift interval \citep[see e.g. Fig. 11 in][]{Despali2016}. Conversely, the SBARBINE suite includes simulation runs with cosmological parameter values sufficiently far from the concordance $\Lambda$CDM model to better probe the cosmological dependence of the mass function on $\Omega_m$ and $\sigma_8$ than the ST-RayGal case. Therefore, this suggests that a simulation suite consisting of runs with volume and mass resolution similar to that of the RayGalGroupSims simulation for very different cosmological parameter values should provide a mass function parametrisation sufficiently accurate to guarantee unbiased sparsity analyses in the case of sparsity data with errors at $\sim 1\%$ statistical level.

\subsection{Hydrostatic Mass Estimates}\label{barsys}
Numerical simulation studies \citep[see e.g.][]{Nagai2007,Meneghetti2010,Rasia2012,Velliscig2014,Biffi2016} as well as the analyses of observed cluster samples \citep{SerenoEttori2015} have shown that X-ray cluster masses obtained by solving the hydrostatic equilibrium (HE) equation are systematically underestimated compared to the true mass of the clusters.

The halo sparsity is unaltered by a constant systematic mass bias, since it is a mass ratio. In contrast, a radial dependent shift affecting HE masses can alter the sparsity\footnote{Let be $M_{\Delta}^t$ the true halo mass and $M_{\Delta}^e$ the estimated one at overdensity $\Delta$ respectively. We define the fraction mass bias as $y_{\Delta}\equiv (M_{\Delta}^e-M_{\Delta}^t)/M_{\Delta}^t$. Then, the relative variation of the halo sparsity compared to its true value is given by
\begin{equation}\label{sparsbias}
r_{\Delta_1,\Delta_2}\equiv\frac{\Delta s_{\Delta_1,\Delta_2}}{s_{\Delta_1,\Delta_2}}=\frac{1+ y_{\Delta_1}}{1+ y_{\Delta_2}}-1;
\end{equation}
from where it is evident that if the mass bias is independent of the cluster radius, $y_{\Delta_1}=y_{\Delta_2}$ and $\Delta s_{\Delta_1,\Delta_2}=0$.} and introduce a systematic error in the cosmological parameter inference. In addition to a bias effect, HE masses suffer from an intrinsic scatter of order of $10-20\%$ \citep{Rasia2012,SerenoEttori2015}. However, most of the sources of scatter act similarly over different radial ranges, so that this would induce negligible effects on sparsity.

The overall amplitude of the radial dependent mass bias has been estimated in several numerical simulation studies. Nevertheless, the results differ as consequence of the different numerical schemes used in the realisation of the simulations as well as to the modelling and the implementation of the astrophysical process that shape the properties of the gas in clusters. As an example \citet{Rasia2012} have realised zoom simulations of $20$ clusters at $z=0.2$ with $M_{200c}>4\cdot10^{14}$ M$_\odot$ h$^{-1}$ and found a $33\%$ median mass bias at $r_{500c}$ and $27.5\%$ at $r_{1000c}$ \citep[see Table 2 in][]{Rasia2012}. These induce a relative shift with respect to the true average sparsity of $\sim 8\%$. A smaller amplitude of the mass bias has been found by \citet{Nagai2007}, nevertheless both these studies have neglected the impact of active galactic nuclei (AGN) on the halo mass.

\begin{figure}[t]
\centering
\includegraphics[width=0.45\textwidth]{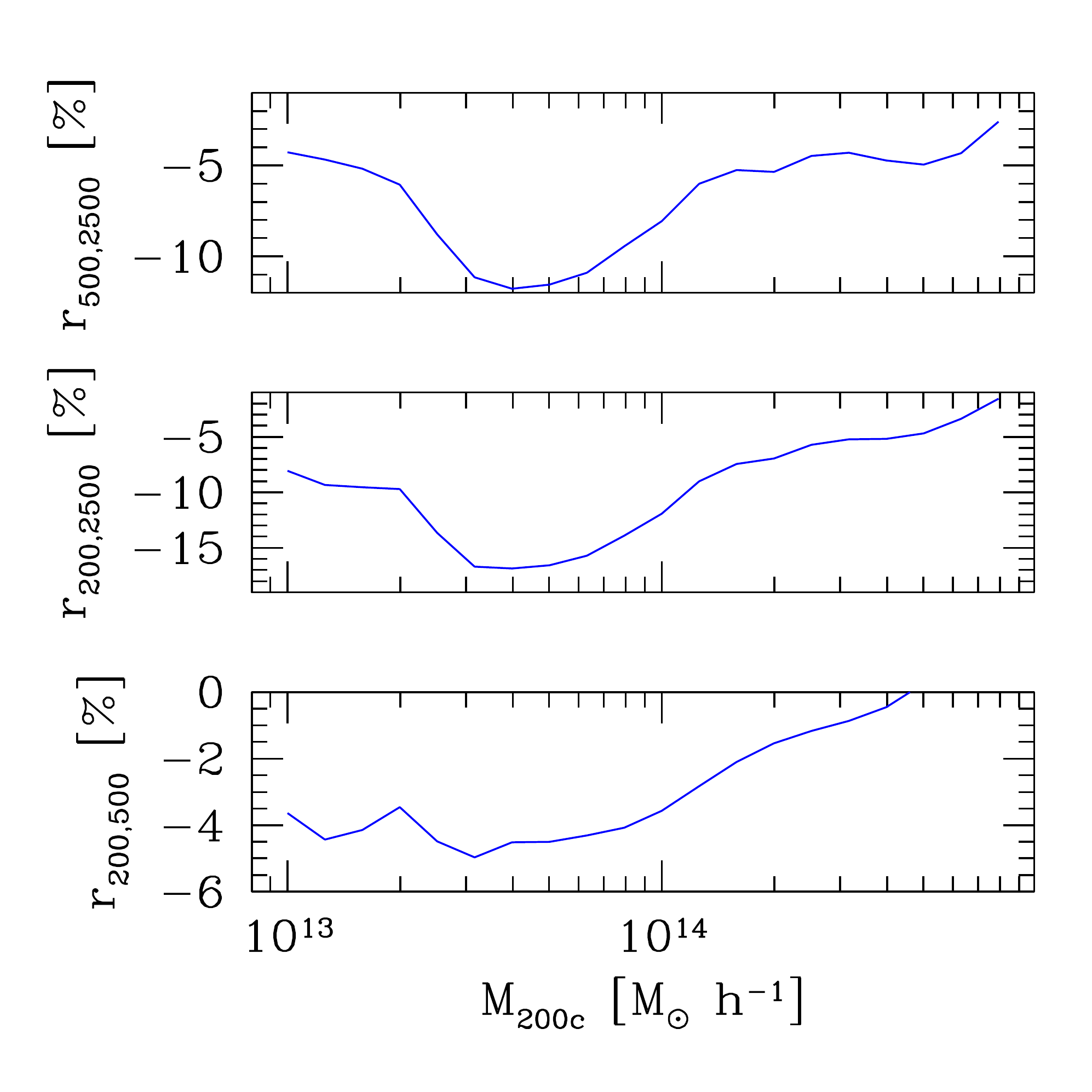}
\caption{Percentage variation of the median halo sparsity $s_{200,500}$ (bottom panel), $s_{200,2500}$ (middle panel) and $s_{500,2500}$ (top panel) due to the radial mass bias induced by baryonic feedback processes as in the AGN 8.0 model investigated in \citep{Velliscig2014}.}\label{agnbias}
\end{figure}

The OverWhelmingly Large Simulations (OWLS) project \citep{Schaye2010} has performed a comprehensive study of the impact of baryonic feedback processes such as star formation, metal-line cooling, stellar winds, supernovae and AGN on the properties of galaxy clusters. Quite remarkably these simulations reproduce the optical and X-ray observed features of groups and clusters of galaxies \citep{McCarthy2010,LeBrun2014}. The effects induced on the halo mass have been studied in detail in \citet{Velliscig2014}. In this study, the authors have evaluated the median fractional mass bias $y_{\Delta}$ at $z=0$ for $\Delta=200c,500c$ and $2500c$ as function of the halo DM mass. Their results have shown that baryonic effects can alter the total halo mass at $\sim 15-20\%$ level for halos with $M_{200c}\sim 10^{13}$ M$_{\odot}$ h$^{-1}$ down to few percent for the most massive systems with $M_{200c}\sim 10^{15}$ M$_{\odot}$ h$^{-1}$. We use their results for the feedback model AGN 8.0 \citep[see Fig.~2 in][]{Velliscig2014} reproducing the observed X-ray profile of clusters \citep{LeBrun2014}. In Fig.~\ref{agnbias} we plot the percentage variation of the median halo sparsity for $s_{200,500}$ (bottom panel), $s_{200,2500}$ (middle panel) and $s_{500,2500}$ (top panel) as given by Eq.~(\ref{sparsbias}). We can see that at large radii the effect of baryonic feedback cause the sparsity to be understimated by $\Delta\langle s_{200,500}\rangle/\langle s_{200,500}\rangle \lesssim 4\%$. For inner radii the effect is larger, but limited $\lesssim 15\%$ for $s_{200,2500}$ and $\lesssim 10\%$ for $s_{500,2500}$. In any case, we notice that for massive systems with $M_{200c}\gtrsim 10^{14}$ M$_{\odot}$ h$^{-1}$ the level of bias on the sparsity is below $\sim 5\%$.

The study presented in \citet{Velliscig2014} has focused on how baryonic processes alter halo masses. On the other hand, in our analysis we are particularly interested on the effects on the HE estimated masses. This has been recently investigated by \citet{Biffi2016}, who have performed zoom simulations of 29 clusters at $z=0$ with masses ${\rm M}_{200c}\gtrsim 10^{14}$ M$_{\odot}$ h$^{-1}$. These simulations account for metallicity-dependent radiative cooling, time-dependent UV background, star formation, metal enrichment, stellar winds and AGN feedback. The authors have estimated the fractional median hydrostatic mass bias for cool-core (CC), non-CC, regular and disturbed systems for overdensity thresholds $\Delta=200c,500c$ and $2500c$ \citep[see Table 1 in ][]{Biffi2016}. Using these results we linearly extrapolate the hydrostatic mass bias at $\Delta=1000c$ and compute the fraction bias on the sparsity $s_{500,1000}$ which we report in Table~\ref{table_bias} for different cluster categories. We can see that they hydrostatic mass bias induce a shift on the true cluster sparsity $0.1-0.3\%$ (non-CC and regular) and $2-4\%$ (CC and disturbed), which is largely in agreement with the estimates we have obtained using the results from \citet{Velliscig2014}.

\begin{table}[t]
\centering
\begin{tabular}{|c|c|}
\hline
Cluster State & $r_{500,1000}$ [\%] \\
\hline
CC &  $-3.7$\\
\hline
NCC & $0.1$\\
\hline
Regular & $0.3$\\
\hline
Disturbed & $-2.0$\\
\hline
\end{tabular}
\caption{Relative variation of the sparsity $s_{500,1000}$ due to hydrostatic mass bias for CC, NCC, regular and disturbed clusters simulated in \citep{Biffi2016}.}\label{table_bias}
\end{table}

Further numerical analyses are nonetheless necessary since no study has so far investigated in detail how the hydrostatic mass bias evolves with time and therefore how the bias on the sparsity evolves with redshift. \citet{Velliscig2014} have shown that the baryonic effects that alter $M_{200c}$ at $z=0$ tends to be smaller (by $\sim 5\%$) at $z=1$. If such a trend holds for larger overdensity thresholds, that would imply that the bias on the halo sparsity is a decreasing function of redshift.
  
Overall, all these elements confirm the strength of the cluster sparsity against possible mass bias systematics. The advantage is twofold. In fact, being a mass ratio, any systematic error affecting cluster mass estimates is suppressed. Moreover, one can focus on the sparsity at overdensity thresholds corresponding to external regions of the cluster mass profile where baryonic effects are subdominant. It is also worth noticing that though hydrostatic masses depend on the choice of a fiducial cosmology through the angular diameter distance, the sparsity, being a mass ratio, is independent of such an assumption.

\begin{figure}[ht]
\centering
\includegraphics[width=0.45\textwidth]{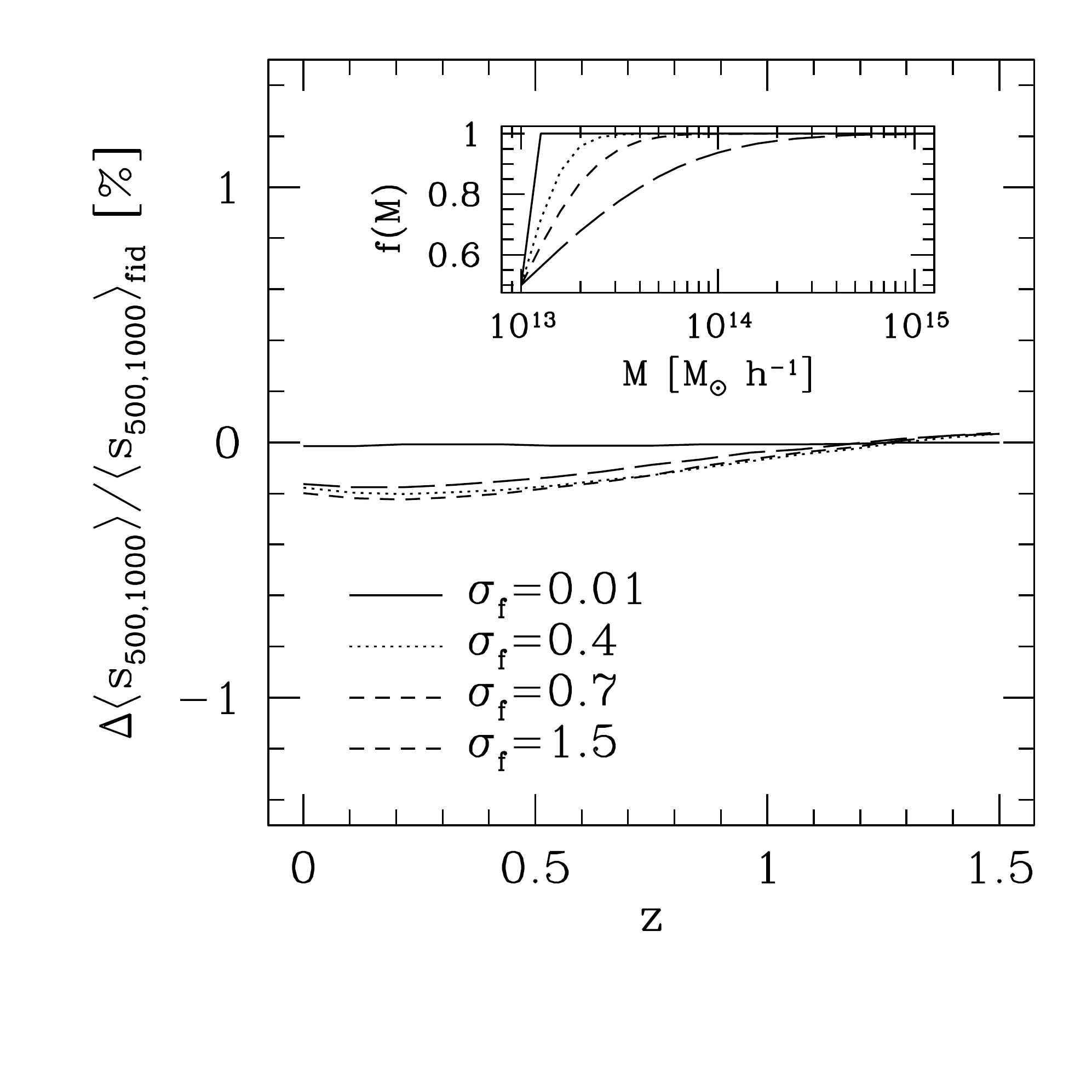}
\caption{\label{fig9} Relative difference of the average sparsity with respect to the case with no selection function for the {\it Planck} fiducial cosmology and different values of the selection function parameter $\sigma_f=0.01,0.4,0.7$ and $1.5$. The inset plot shows the form of the selection function for the different values of $\sigma_f$.}
\end{figure}

\subsection{Selection Effects}
A final remark concerns selection effects. In principle we do not expect a significant contribution since we have seen that average sparsity as predicted by Eq.~(\ref{sparpred}) is largely independent of the lower limit of integration. To have a quantitative estimate of potential systematics induced by the shape of the selection function, we multiply the integrands on both sides of Eq.~(\ref{sparpred}) by a selection function of the form:
\begin{equation}
f({\rm M}_{\Delta})=\frac{1}{2}\left[1+{\rm erf}\left(\frac{\ln{{\rm M}_{\Delta}}-\ln{{\rm M}^{\rm min}_{\Delta}}}{\sqrt{2}\sigma_f}\right)\right],
\end{equation}
where $\sigma_f$ modulate the shape of the selection function. 

In Fig.~\ref{fig9} we plot the relative difference of the redshift evolution of the average sparsity with respect to the case $f({\rm M}_{\Delta})=1$ for the fiducial {\it Planck} cosmology and for different values of $\sigma_f=0.01,0.4,0.7$ and $1.5$. We can see the differences are at sub-percent level.

\section{Sparsity of X-ray Clusters \& Cosmological Parameter Constraints}\label{data}

\begin{figure}[t]
\centering
\includegraphics[width=.45\textwidth]{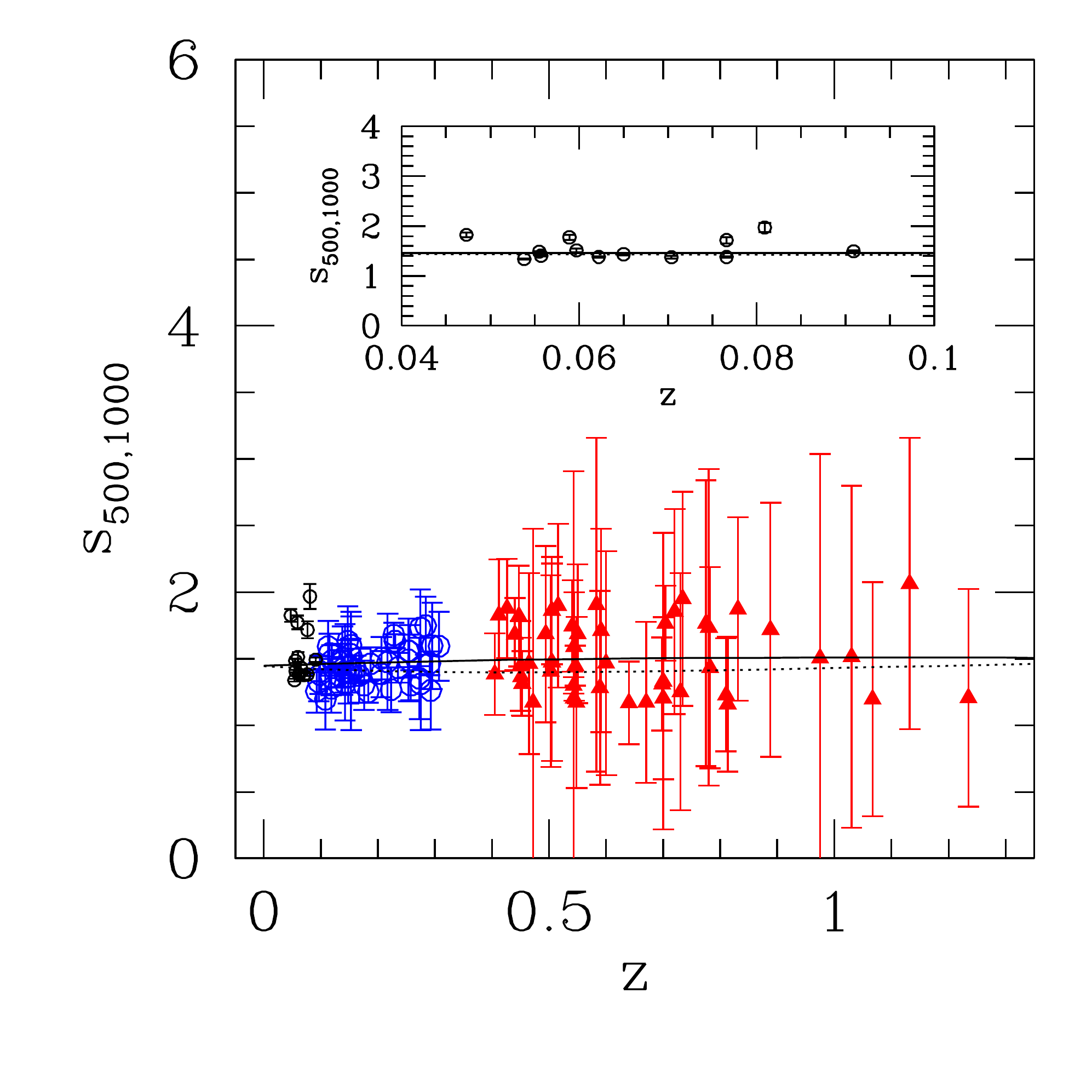}
\caption{\label{fig5} Sparsity of X-ray clusters. The low-redshift sample consists of clusters with mass measurements from \citet{Ettori2017,Ettori2017b} and \citet{Ghirardini2017} also shown in the inset plot (black empty circles), and \citet{Ettori2010} (blue empty circles). The high-redshift sample consists of clusters with mass estimates from \citet{Amodeo2016} (red solid triangles).  The black solid line and the black dotted line correspond to the best-fit $\Lambda$CDM models inferred assuming the ST-RayGal and ST-Despali mass functions respectively. We may notice four clusters at $z<0.1$ whose sparsity significantly depart from the best-fit. We have checked that excluding these outliers from the analysis does not alter the result of the cosmological parameter inference.}
\end{figure}

We estimate the halo sparsity of a set of X-ray galaxy clusters with hydrostatic mass measurements. The dataset consists of a low-redshift sample of 57 clusters ($0.05<z<0.3$) from \citet{Ettori2010,Ettori2017,Ettori2017b} and \citet{Ghirardini2017} and a high-redshift sample of 47 clusters ($0.4<z<1.2$) presented in \citet{Amodeo2016}. DM masses ${\rm M}_{500c}$ and ${\rm M}_{1000c}$ have been estimated by solving the hydrostatic equilibrium equation \cite[see e.g.][]{Sarazin1986,Ettori2013}. We compute the sparsity of each cluster in the catalogs, $\hat{s}_{500,1000}={\rm M}_{500c}/{\rm M}_{1000c}$, and estimate the uncertainty by propagating the mass measurement errors. These are shown in Fig.~\ref{fig5}. 

For simplicity we have neglected mass correlation effects: these may be present due to the mass measurement methodology which assumes a functional form of the DM halo profile\footnote{Given that the sparsity is a mass ratio, a positive correlation $r$ between the estimates of $M_{500c}$ and $M_{1000c}$ would imply that we are overestimating the sparsity errors by a factor $\sim 1/\sqrt(1-r)$. For instance, if $r\sim 0.5$ as it is reasonable to expect, this would correspond to a $30\%$ overestimation and thus result into more conservative constraints on the cosmological parameters.}. Systematics affecting the HE mass estimate can be more important. In the case of the high-redshift sample, \citet{Amodeo2016} have tested the consistency of the HE masses for a subset of 32 clusters for which gravitational lensing measurements  were available in the literature \citep[{\it LC}$^2$-{\tt single} catalogue from][]{Sereno2015}. They have found a good agreement within the large statistical uncertainties with $\ln{({\rm M}_{HE}/{\rm M}_{\rm lens})}=0.16\pm 0.65$. For the low-redshift sample there is no available comparison, however we noticed that the dataset from \citet{Ettori2010} consists of massive clusters for which mass measurement errors are larger than the expected bias from baryonic feedback discussed in Section~\ref{barsys}. In the case of the very low-redshift sample by \citet{Ettori2017,Ettori2017b,Ghirardini2017}, HE mass uncertainties are at a few percent level and we cannot a priori exclude that some of the sparsity measurements are affected by radial mass bias. After all, we can see in Fig.~\ref{fig5} that the sparsity of four of the clusters in the very low-redshift sample significantly deviate from the values of the other objects in the dataset. We have found that removing these objects from the data analysis leaves the cosmological results unaltered. As seen in Section~\ref{cosmodep}, this is direct consequence of the fact that the cosmological signal is largest at $z\sim 0.5$. Nevertheless, to test the stability of the cosmological analysis against possible contamination from HE mass bias, we perform an additional analysis assuming a systematic redshift dependent shift of the measured cluster sparsity. More specifically, we assume a $5\%$ shift of the cluster sparsity at $z=0$ linearly reducing to $2\%$ at $z=1$. This is an extremely conservative bias model especially if compared to the level of bias discussed in Section~\ref{barsys}. In fact, it implies that the HE mass determination of each cluster systematically underestimates $M_{500c}$ by $28\%$ and $M_{1000c}$ by $35\%$ at $z=0$, and by $23\%$ and $30\%$ respectively at $z=1$ \citep[consistently with the $5\%$ reduction found for $M_{200c}$ estimates in][]{Velliscig2014}.

Since we compare individual cluster sparsity estimates to the predictions of the ensemble average sparsity we account for the intrinsic dispersion of the halo sparsity discussed in Section~\ref{nbodytest} by adding in quadrature a conservative and absolute $0.2$ intrinsic scatter (consistent with N-body simulation results shown in the inset plot of Fig.~\ref{fig:sparsz_th_vs_nh}) to the statistical error. 

We perform a Markov Chain Monte Carlo data analysis to derive constraints on the $\Lambda$CDM model parameters, $(\Omega_m,\sigma_8,h,n_s,\Omega_b h^2)$. To reduce the effect of parameter degeneracies, we assume a set of Gaussian priors on $n_s\sim \mathcal{N}(0.963,009)$ consistently with {\it Planck} results \citep{PlanckCosmo}, $h\sim \mathcal{N}(0.688,0.033)$ from \citet{Efstathiou2014} and $\Omega_b h^2\sim \mathcal{N}(0.022,0.002)$ consistent with Big-Bang Nucleosynthesis bounds \citep{BBN}. We assume flat priors for $\Omega_m\sim \mathcal{U}(0.05,0.95)$ and $\sigma_8\sim \mathcal{U}(0.2,1.8)$. In order to evaluate the impact of the prior on $h$, we have also performed a likelihood analysis of the full cluster sample assuming a Gaussian HST prior $h\sim \mathcal{N}(0.732,0.024)$ from \citet{Riess2016}.

We evaluate the following $\chi^2$:
\begin{equation}
\chi^2=\sum_i \frac{[\hat{s}^{i}_{500,1000}-\langle s^{\rm th}_{500,1000}(z_i)\rangle]^2}{\sigma_{\rm int}^2+\sigma^2_{s^i_{500,1000}}},
\end{equation}
where $\hat{s}^{i}_{500,1000}$ is the sparsity of the $i$-th cluster in the catalog, $\langle s^{\rm th}_{500,1000}(z_i)\rangle$ is the theoretical model prediction given by Eq.~(\ref{sparpred}) assuming a given mass function model, $\sigma_{\rm int}=0.2$ is the intrinsic scatter of the halo sparsity\footnote{In principle one can attempt to infer the intrinsic scatter from the data regression with the other cosmological parameters.} (conservatively set to a value consistent with the N-body results) and $\sigma_{s^i_{500,1000}}$ the error on the cluster sparsity measurement. 

We use the Metropolis-Hastings algorithm to generate 15 independent random chains of $2\times 10^5$ samples and evaluate the rejection rate every 100 steps and adjust the width of the parameters dynamically. We check the convergence of the chains using the Gelman-Rubin test \citep{Gelman}.

\begin{figure}[th]
\centering
\includegraphics[width=0.45\textwidth]{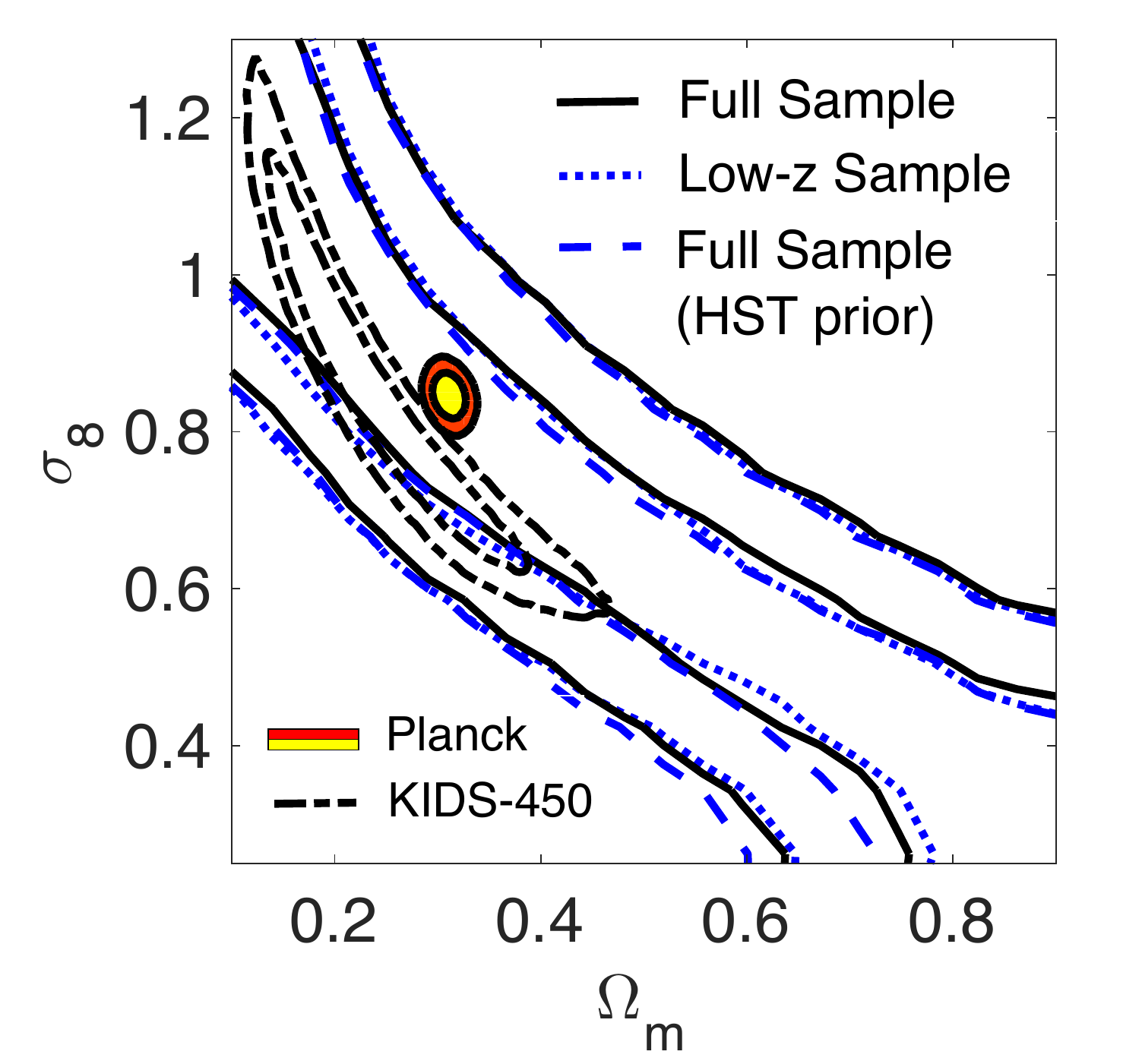}
\caption{\label{fig6} Marginalised $1$ and $2\sigma$ contours in the $\Omega_m-\sigma_8$ plane assuming the ST-RayGal mass function using the full X-ray cluster dataset (black solid lines), the low-z redshift sample only (blue dotted lines) and in the case of the HST prior on $h$ (blue dashed lines). For comparison we also plot the contours from the {\it Planck} cosmological data analysis \citep{PlanckCosmo} and KIDS-450 \citep{KIDS}.}
\end{figure}

\begin{figure}[th]
\centering
\includegraphics[width=0.45\textwidth]{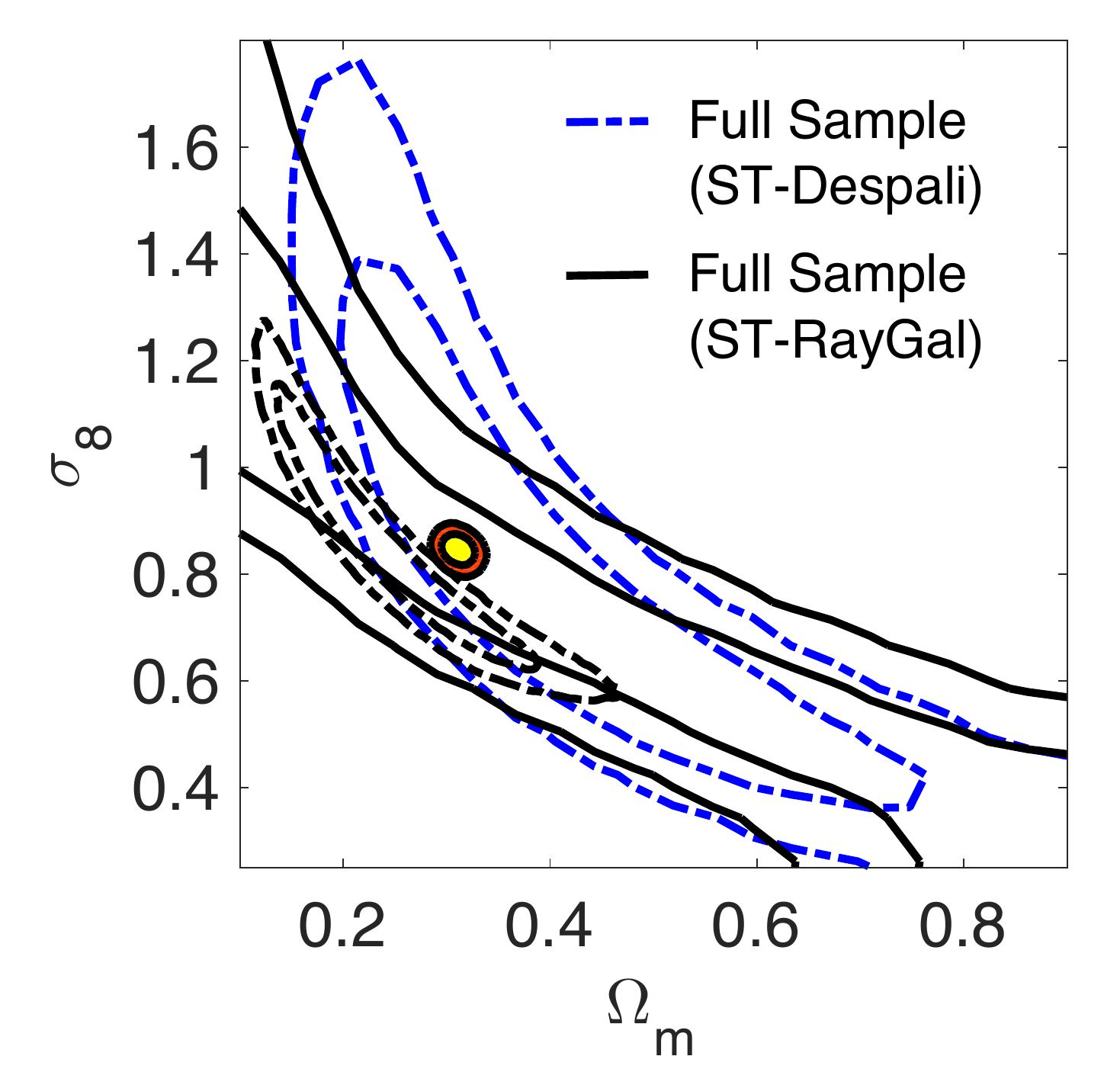}
\caption{\label{fig6bis} Marginalised $1$ and $2\sigma$ contours in the $\Omega_m-\sigma_8$ plane as in Fig.~\ref{fig6}. Here, the blue long-short dashed lines are the contours inferred assuming the ST-Despali mass function.}
\end{figure}

The results of the likelihood data analysis assuming the ST-RayGal mass function are summarised in Fig.~\ref{fig6} where we plot the marginalised $1$ and $2\sigma$ credibility contours in the $\Omega_m-\sigma_8$ plane for the different cases. For comparison we also plot the marginalised credibility contours from the {\it Planck} cosmological data analysis \citep{PlanckCosmo} and the weak gravitational lensing from KIDS-450 \citep{KIDS}. We can see that the constraints on $\Omega_m$ and $\sigma_8$ are rather weak. Given the large uncertainties of the sparsity sample at $z\gtrsim 0.4$, this is not surprising since the variation of the sparsity with respect to $\sigma_8$ is the largest at $z\sim 0.4$, while that with respect to $\Omega_m$ remains quite flat for $z\gtrsim 0.5$ (see top panel in Fig~\ref{fig:cosmo_derivs}). The best-fit model corresponds to $\Omega_m\simeq 0.4$ and $\sigma_8\simeq 0.6$. We plot the associated average sparsity as function of redshift as black solid line in Fig.~\ref{fig5}. Notice the strong degeneracy between $\Omega_m$ and $\sigma_8$. As discussed in Section~\ref{cosmodep} this is expected given the sensitivity of the halo sparsity to $S_8\equiv \sigma_8\sqrt{\Omega_m}$ for which we find $S_8=0.40\pm 0.11$ at $1\sigma$.

As we can see in Fig.~\ref{fig6}, the credibility contours do not significantly differ from those inferred under the HST prior. This is also consistent with the analysis presented in Section~\ref{cosmodep}, which indicates that the halo sparsity is less sensitive to $h$ than $\sigma_8$, $\Omega_m$ and $n_s$. Indeed, changing the priors on $n_s$ can have a more significant impact on the inferred constraints. However, $n_s$ is tightly constrained by the {\it Planck} data, while there are larger uncertainties on the value of $h$, that is why we have tested the sensitivity of the constraints to the $h$ prior.

In Fig.~\ref{fig6} we also plot the credibility contours inferred using the low-z sample ($z\lesssim 0.4$) only. These do not differ from those obtained using the full sample, which is not surprising given the larger uncertainties of the high-z sample. Overall, the inferred credibility contours overlap with those inferred from {\it Planck} within $1\sigma$ as well as those from the KIDS-450 dataset.

In Fig.~\ref{fig6bis} we plot the constraints in the $\Omega_m-\sigma_8$ plane inferred assuming the ST-Despali mass function. Differently from the ST-RayGal case we find bounded contours at $1\sigma$, though still spread over a larger portion of the parameter space. The one-dimensional marginalised constraints are $\Omega_m=0.42\pm 0.17$ and $\sigma_8=0.80\pm 0.31$ at $1\sigma$, with the best-fit values being $\Omega_m=0.36$ and $\sigma_8=0.74$. We plot the associated average sparsity as function of redshift as black dotted line in Fig.~\ref{fig5}. From the analysis of the Monte Carlo chains we obtain $S_8=0.48\pm 11$ at $1\sigma$, which is consistent with the constraints found using the ST-RayGal mass function. As we can see in  Fig.~\ref{fig6bis}, the contours are statistically consistent with those inferred from the ST-RayGal analysis, though deviations are noticeable in the tail of the distribution for low values of $\Omega_m$ and large values of $\sigma_8$. This is not unexpected since in this range of the parameter space the mass function calibration may deviate from that of the vanilla $\Lambda$CDM model of the RayGalGroupSims simulation. The bounds are compatible with the {\it Planck} results and consistent with those from the KIDS-450 analysis\footnote{Several large-scale structure data analyses have constrained combinations of $\Omega_m$ and $\sigma_8$. As an example, SZ cluster abundance data from the South Pole Telescope (SPT) survey gives $\sigma_8(\Omega_m/0.27)^{0.3}=0.797\pm 0.031$ \citep{Haan2016}. The analysis of the cluster sparsity presented here gives consistent bounds, $\sigma_8(\Omega_m/0.27)^{0.3}=0.87\pm 0.26$. Similarly, measurements of the galaxy clustering from the Dark Energy Survey (DES) constrain $\sigma_8(\Omega_m/0.3)^{0.16}=0.74\pm 0.12$ \citep{Kwan2016}, and we find again a result consistent within $1\sigma$, $\sigma_8(\Omega_m/0.3)^{0.16}=0.83\pm 0.29$.}.

We have limited the analysis including the systematic HE mass bias model discussed at the beginning of this section to the case of the ST-Despali mass function. The results of the likelihood data analysis give $S_8=0.51\pm 0.11$, which is consistent with the results obtained assuming no systematic bias model.

\section{X-ray Cluster Sparsity Forecasts}\label{forecast}
Future observational programs will provide increasingly large samples of clusters. Surveys such as {\it eROSITA} \citep{eRositabook} are expected to detect several hundred thousands of clusters across a large redshift range. Cosmological parameter constraints will be inferred from accurate measurements of cluster number counts and spatial clustering \citep[see e.g.][]{Pillepich2012}. 

Sparsity measurements capable of providing constraints that are competitive with respect to those inferred from other cosmological probes strongly depends on the availability of accurate mass estimations. In the case of large datasets, such as those from {\it eROSITA}, cluster masses will be measured through the use of observationally calibrated scaling relations \citep[see e.g.][]{Maughan2012,Ettori2013,Ettori2015}. More precise estimates for instance using HE masses require observations that are able to resolve the cluster mass profile. However, these may be available only for smaller cluster samples through follow-up observations. 

Here, we perform a Fisher matrix forecast of the cosmological parameter errors from sparsity measurements to determine the type of galaxy cluster observations needed to derive competitive constraints with respect to those that can be obtained with other standard probes such as the CMB.

To this purpose we evaluate the Fisher matrix:
\begin{equation}
F_{\mu\nu}=\sum_i \frac{1}{\sigma_{z_i}^2}\frac{\partial \langle s_{500,1000}(z_i)\rangle}{\partial\theta_{\mu}}\frac{\partial \langle s_{500,1000}(z_i)\rangle}{\partial\theta_{\nu}}\biggr\rvert_{\hat{\theta}_{\mu}},\label{fisherfore}
\end{equation}
where $\theta_{\mu}=(\Omega_m,\sigma_8,h,n_s,\Omega_b)$ are the cosmological parameters, $\hat{\theta}_{\mu}$ the fiducial parameter values and $\sigma_{z_i}$ is the statistical error on the mean sparsity. We compute the partial derivatives in Eq.~(\ref{fisherfore}) using a five-point stencil approximation. We model the error on the average sparsity as 
\begin{equation}
\sigma_{z_i}=\langle s^{\rm fid}_{500,1000}(z_i)\rangle\,e_{M}\sqrt{\frac{2}{N(z_i)}},
\end{equation}
where $\langle s^{\rm fid}_{500,1000}(z_i)\rangle$ is the fiducial sparsity value, $e_{M}$ is the fraction error on mass measurements, $N(z_i)$ is the number of clusters at redshift $z_i$ given by
\begin{equation}
N(z_i)\equiv A_{\rm survey}\,\Delta{z}\,f\,\frac{dN}{dzdA}(z_i),
\end{equation}
where $A_{\rm survey}$ is the survey area, $\Delta{z}$ is the size of the redshift bins, $f$ is the fraction of clusters with mass measurement error $e_M$ and $\frac{dN}{dzdA}(z)$ is cluster number count distribution. Again for simplicity we neglect correlations in the estimation of the masses ${\rm M}_{500c}$ and ${\rm M}_{1000c}$. Notice that as in the case of the synthetic likelihood test presented in Section~\ref{data}, we do not add in quadrature the intrinsic scatter of the halo sparsity to the statistical error as in the data analysis described in Section~\ref{data}. This is because in the spirit of the Fisher matrix calculation we do not compare predictions of the ensemble average sparsity to the sparsity of an individual cluster at a given redshift, rather to the estimated average sparsity from an ensemble of $N(z_i)$ clusters.

\begin{figure}[t]
\centering
\includegraphics[width=0.47\textwidth]{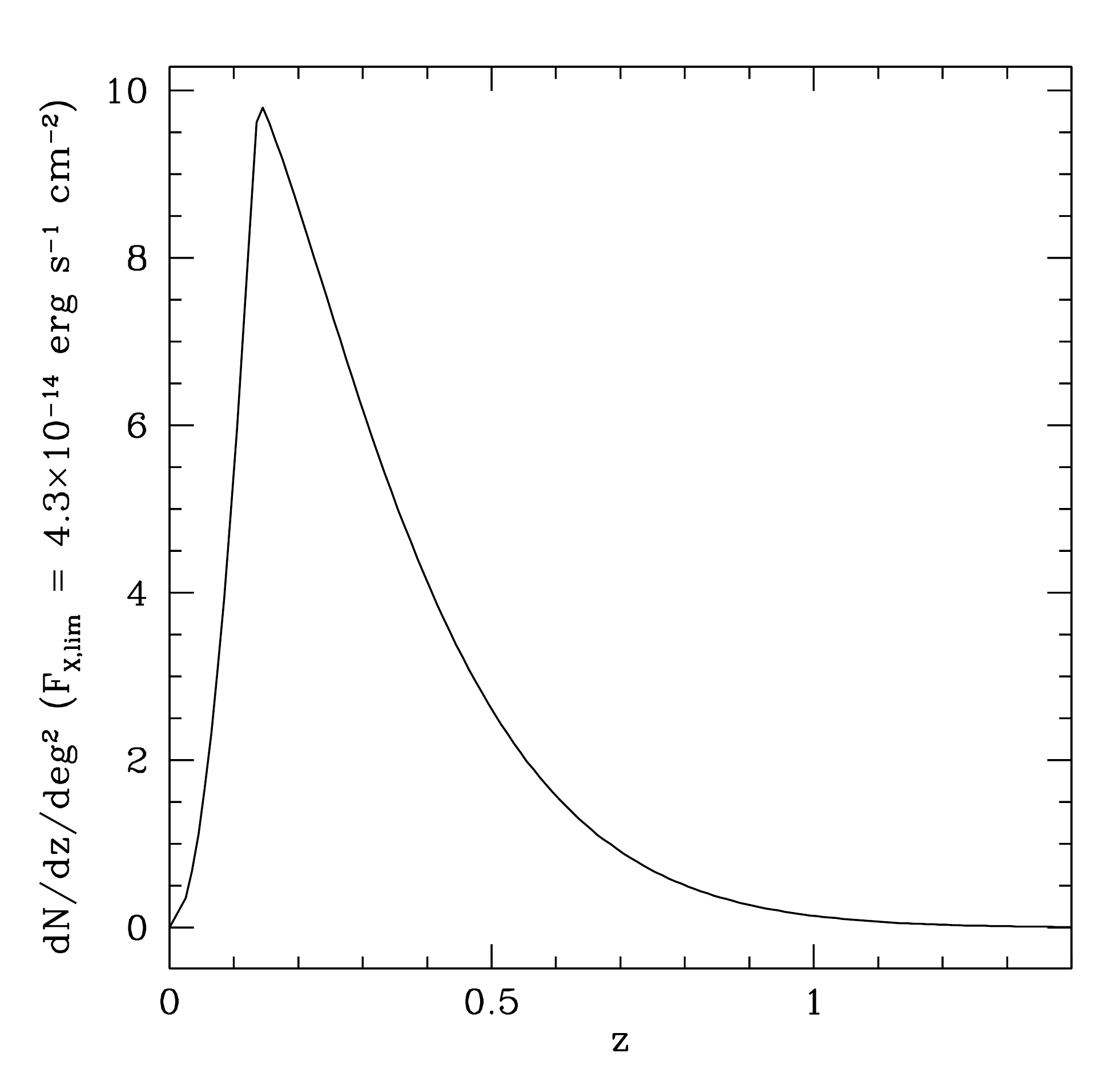}
\caption{\label{erositafig} Expected redshift distribution of clusters of a {\it eROSITA}-like survey with X-ray flux cut $F_{\rm X,cut}=4.3\times 10^{-14}$ erg s$^{-1}$ cm$^{-2}$ for our fiducial cosmological model.}
\end{figure}

We assume a {\it Planck} fiducial $\Lambda$CDM cosmology and consider a full sky survey with cluster number count distribution consistent with a {\it eROSITA}-like survey. This is expected to detect $\sim 10^5$ clusters with mass $\gtrsim 10^{13}$ h$^{-1}$ M$_{\odot}$. To this purpose we estimate the cluster number counts as function of redshift for our fiducial cosmology by integrating the ST-RayGal mass function with ${\rm M}_{500c}$ and imposing a flux cut $F_{\rm X,cut}=4.3\times 10^{-14}$ erg s$^{-1}$ cm$^{-2}$, where we have used the luminosity-mass relation from \citet{Mantz2010} with no intrinsic scatter. The predicted number count distribution is shown in Fig.~\ref{erositafig}. We may notice that this is consistent with the redshift distribution estimated by \citet{Pillepich2012} (see their Fig.~3 for the photon count rate threshold corresponding to ${\rm M}^{\rm cut}_{500c}\gtrsim 5\times 10^{13}$ h$^{-1}$ M$_{\odot}$) with a total count of $\sim 3\times 10^4$ clusters. For simplicity, here we only consider redshift bins of size $\Delta{z}=0.1$ in the redshift range $0\la z\la 1.4$. 
 
We limit our analysis to two different observational scenarios: {\it small sample-high precision} sparsity measurements with mass errors of $e_M=0.01$  for $\sim 300$ clusters ($f=0.01$) and $e_M=0.05$ for $\sim 3000$ clusters ($f=0.10$) respectively; {\it large sample-low precision} sparsity measurements with mass errors of $e_M=0.1$ for $\sim 6000$ clusters ($f=0.2$) and $e_M=0.2$ for all clusters ($f=1$). The latter scenario considers the possibility of measuring masses over a large sample of clusters through well calibrated scaling relations whose validity should be limited to $\sim 10-30\%$ level.

We combine the Fisher matrix from Eq.~(\ref{fisherfore}) to the {\it Planck}-Fisher matrix which has been computed using the code CosmoFish \citep{CosmoFish1,CosmoFish2}. 

The results are summarised in Tables~\ref{tab_ideal} and~\ref{tab_real}. In all cases we can see that including the information from the halo sparsity improves the CMB constraints from {\it Planck}. Indeed, the level of improvement depends on the observational configuration considered. Quite remarkably, we find that a $1\%$ mass error estimation for a sample of $\sim 300$ clusters has the greatest impact in reducing the {\it Planck} errors on several parameters. For instance, we find an improvement of a factor $\sim 2.3$ on the estimation of $\sigma_{\sigma_8}$ in the ST-RayGal case, while assuming the ST-Despali mass function we find an improvement of a factor $\sim 1.9$ on $\sigma_{\Omega_m}$ and $\sim 1.8$ on $\sigma_{\sigma_8}$. Even for the realistic scenario with $20\%$ mass errors, we find up to $\sim 30\%$ improvement of the {\it Planck} constraints. Again, assuming the ST-Despali mass function systematically predicts smaller parameter errors than those obtained the ST-RayGal mass function.  

Compared to other cosmic probes such as the combination of CMB constraints with those from cluster number counts and angular clustering studied in \citet{Pillepich2012} we find that the sparsity can provide cosmological parameter constraints of the same order  \citep[see e.g. Table B2 in][]{Pillepich2012}.

\begin{table*}[t]
\centering
\begin{tabular}{|c|c|c|c|}
\hline
  & {\it Planck} only & $+\langle s_{500,1000}(z)\rangle$ ($e_M=0.01$, $f=0.01$)& $+\langle s_{500,1000}(z) \rangle$ ($e_M=0.05$, $f=0.1$) \\
\hline
 $\sigma_{\Omega_m}$ &  $0.01082$ & $0.01022$ / $0.00571$ & $0.01035$ / $0.00684$\\
\hline
 $\sigma_{\sigma_8}$ &  $0.01396$ & $0.00597$ / $0.00780$ & $0.00720$ / $0.00926$ \\
\hline
$\sigma_{n_s}$ & $0.00428$ & $0.00414$ / $0.00300$ & $0.00418$ / $0.00328$ \\
\hline
$\sigma_h$ & $0.00763$ & $0.00723$ / $0.00411$  & $0.00732$ / $0.00490$\\
\hline
$\sigma_{\Omega_b}$ & $0.00095$ & $0.00090$ / $0.00053$  & $0.00091$ / $0.00062$ \\
\hline
\end{tabular}
\caption{Marginalised errors on cosmological parameters from the Fisher matrix analysis of {\it small sample-high precision} sparsity measurements in combination with {\it Planck} constraints. The numbers quoted on the left (right) correspond to the Fisher forecast based on the ST-RayGal (ST-Despali) mass function.}\label{tab_ideal}
\end{table*}

\begin{table*}[t]
\centering
\begin{tabular}{|c|c|c|c|}
\hline
  & {\it Planck} only & $+\langle s_{500,1000}(z)\rangle$ ($e_M=0.10$, $f=0.2$) & $+\langle s_{500,1000}(z)\rangle$ ($e_M=0.20$, $f=1$)\\
\hline
 $\sigma_{\Omega_m}$ &  $0.01082$ & $0.01044$ / $0.00751$ & $0.01041$ / $0.00731$ \\
\hline
 $\sigma_{\sigma_8}$ &  $0.01396$ & $0.00850$ / $0.01010$& $0.00805$ / $0.00984$\\
\hline
$\sigma_{n_s}$ & $0.00428$ & $0.00420$ / $0.00345$& $0.00419$ / $0.00340$\\
\hline
$\sigma_h$ & $0.00763$ & $0.00738$ / $0.00536$& $0.00736$ / $0.00522$\\
\hline
$\sigma_{\Omega_b}$ & $0.00095$ & $0.00092$ / $0.00068$ & $0.00092$ / $0.00066$\\
\hline
\end{tabular}
\caption{As in Table~\ref{tab_ideal} for {\it large sample-low precision} sparsity measurements.}\label{tab_real}
\end{table*}

\section{Conclusions}\label{conclu}
In this work we have presented a first cosmological analysis of the dark matter halo sparsity. This characterises halos in terms of the ratio of halo masses at two different overdensities and carry cosmological information encoded in the mass profile of halos which can be retrieved from mass measurements of galaxy clusters. 

Building upon the work of \citet{Balmes2014} we have tested the sparsity properties using halo catalogs from a large volume high-resolution N-body simulation. In particular, we have shown that the average sparsity of an ensemble of halos can be accurately predicted from prior knowledge of the halo mass function. To this purpose we have introduced the ST-RayGal parameterisation which reproduces to great accuracy the numerical halo mass function for halo masses $M_{200c}$, $M_{500c}$ and $M_{1000c}$, and allows us to recover the measured average sparsity values at different redshift snapshots to sub-percent level. 

We have tested the accuracy of the theoretical predictions assuming other mass function parameterisations proposed in the literature. Depending on the mass function model, we found deviations with respect to the average sparsity from the N-body halo catalogs up to $10\%$ level.

The possibility to predict the average sparsity for a given set of cosmological parameters enables us to perform a cosmological model parameter inference using cluster sparsity measurements. To test this we have generated a synthetic set of data and performed a likelihood analysis from which we have retrieved the input fiducial cosmology.

Systematic errors affecting halo sparsity data analyses may arise primarily from uncertainties in the theoretical modelling of the halo mass function and the radial dependent cluster mass bias from baryonic feedback processes. Here, we have performed an analysis of these systematics. Quite importantly, using results from state-of-art numerical simulations we show that for massive systems baryonic effects alter the halo sparsity at a few percent level. This is a subdominant compared to the uncertainties from mass estimation errors of currently available cluster datasets. We find that cluster selection effects have a negligible impact on sparsity which is an obvious advantage compared to other cluster cosmological proxies such as the number counts or the spatial clustering.

We have estimated the sparsity of a sample of X-ray clusters with hydrostatic mass measurements and performed a Markov Chain Monte Carlo likelihood data analysis to infer constraints $\Omega_m$ and $\sigma_8$. We find weak marginalised bounds on $\Omega_m$ and $\sigma_8$. Assuming the mass function from \citet{Despali2016} gives the strongest bound, in particular we find $\Omega_m=0.42\pm 0.17$ and $\sigma_8=0.80\pm 0.31$ at $1\sigma$, corresponding to $S_8=0.48\pm 0.11$. In all cases the inferred constraints are compatible with those inferred from the {\it Planck} cosmological data analysis within $1\sigma$. We find these results to be stable against a conservative systematic bias model accounting for baryonic effects on cluster mass estimates.

Future cluster surveys can provide larger sparsity datasets. Using a Fisher matrix approach we have investigated their complementarity with respect CMB observations from {\it Planck}. In particular we have performed a parameter error forecast for different observational scenarios and we found that sparsity measurements from a small cluster sample of $\sim 300$ clusters with mass uncertainties of $1\%$ can improve {\it Planck} constraints on $\Omega_m$ and $\sigma_8$ by approximately a factor of $2$. However, this requires a control of systematic errors due to hydrostatic mass bias.

Cluster mass measurements from SZ and lensing observations may also provide viable datasets to estimate the halo sparsity and we leave such studies to future works.

\acknowledgments  
We would like to thank Matteo Martinelli for providing us the {\it Planck}-Fisher matrix. PSC is grateful to Joop Schaye for useful discussions. PSC, YR and MAB are thankful to Fabrice Roy for technical support. SE and MS acknowledge the financial support from contracts ASI-INAF I/009/10/0, NARO15 ASI-INAF I/037/12/0 and ASI 2015-046-R.0. The research leading to these results has received funding from the European Research Council under the European Union Seventh Framework Programme (FP7/2007-2013 Grant Agreement no. 279954). We acknowledge support from the DIM ACAV of the Region \^Ile-de-France. This work was granted access to the HPC resources of TGCC under allocation 2016-042287 made by GENCI (Grand \'Equipement National de Calcul Intensif). 

\bibliographystyle{aasjournal}

\appendix
\section{Halo Mass Function Parametrisation}\label{app_hmf}
We use the numerical mass functions estimated from the RayGalGroupSims simulation SOD halo catalogs with mass ${\rm M}_{200c}$, ${\rm M}_{500c}$ and ${\rm M}_{1000c}$ respectively to calibrate at each redshift snapshot the coefficients of the Sheth-Tormen mass function formula \citep{Sheth1999}: 
\begin{equation}
\frac{dn}{d{\rm M}}=\frac{\rho_m}{\rm M}\left(-\frac{1}{\sigma}\frac{d\sigma}{d{\rm M}}\right)A\frac{\delta_c}{\sigma}\sqrt{\frac{2a}{\pi}}\left[1+\left(a\frac{\delta^2_c}{\sigma^2}\right)^{-p}\right] e^{-\frac{a\delta_c^2}{2\sigma^2}},
\end{equation}
$\rho_m$ is the present mean matter density, $\delta_c$ is the linearly extrapolated spherical collapse threshold which we
compute using the formula from \citet{Kitayama1996} and 
\begin{equation}
\sigma^2({\rm M},z)=\frac{1}{2\pi^2}\int dk\,k^2 P(k,z)\,\tilde{W}^2[k\,R({\rm M})],
\end{equation}
is the variance of linear density field smoothed on a spherical volume of radius $R$ enclosing the mass ${\rm M}=4/3\pi\rho_m R^3$, with $P(k,z)$ being the linear matter power spectrum at redshift $z$ and 
\begin{equation}
\tilde{W}^2[k\,R({\rm M})]=\frac{3}{(kR)^3}[\sin{(kR)}-(kR)\cos{(kR)}].
\end{equation}

\begin{table}[t]
\centering
\begin{tabular}{|c|c|c|c||c|c|c||c|c|c|}
\hline\hline
$z$ & $A_{200c}$ & $a_{200c}$ & $p_{200c}$& $A_{500c}$ & $a_{500c}$ & $p_{500c}$ & $A_{1000c}$ & $a_{1000c}$ & $p_{1000c}$ \\
\hline\hline
0.00 & 0.35884 & 1.2300 & -0.79142 &0.28401 & 1.4568 & -0.67260 &  0.20596 &   1.7978 &  -0.72148 \\
\hline
0.50 & 0.42038 & 0.9752 & -0.54330 & 0.28378  & 1.1859 & -0.58425 & 0.21628  &  1.4134  & -0.53098 \\
\hline
0.66 & 0.35697 & 1.0039 & -0.74000 & 0.27724 &  1.1347 & -0.56833 & 0.21555  &  1.3339 &  -0.47343\\
\hline
1.00 & 0.27751 & 0.9944 & -0.93238 & 0.25082 & 1.0470 & -0.61536 & 0.19365  &  1.2537 &  -0.51639\\
\hline
1.14 & 0.29991 & 0.9505 & -0.83109 & 0.24834 & 1.0429& -0.58941 &  0.19134 &   1.2156 &  -0.48951\\
\hline
1.50 & 0.22855 & 0.9457 & -0.97637 & 0.22936 & 1.0014 & -0.58554 &  0.18885 &   1.1400 &  -0.39898\\
\hline
2.00 & 0.15502 & 0.9375 & -1.13120 & 0.23210 &  0.9459 &  -0.46949 & 0.18072  &  1.0830 &  -0.34148\\
\hline\hline
\end{tabular}\caption{\label{tabA1} Best-fit coefficients of the ST mass function for halos with masses ${\rm M}_{200c}$, ${\rm M}_{500c}$ and ${\rm M}_{1000c}$ respectively.}
\end{table}


We determine the best-fit ST coefficients using a Levenberg-Marquardt minimisation scheme. These are quoted in Table~\ref{tabA1} for halo masses ${\rm M}_{200c}$, ${\rm M}_{500c}$ and ${\rm M}_{1000c}$ respectively. We find the best-fit functions to have logarithmic deviations with respect to the numerical estimates to better than $5\%$.

\begin{figure}[h]
\centering
\includegraphics[width=.4\textwidth]{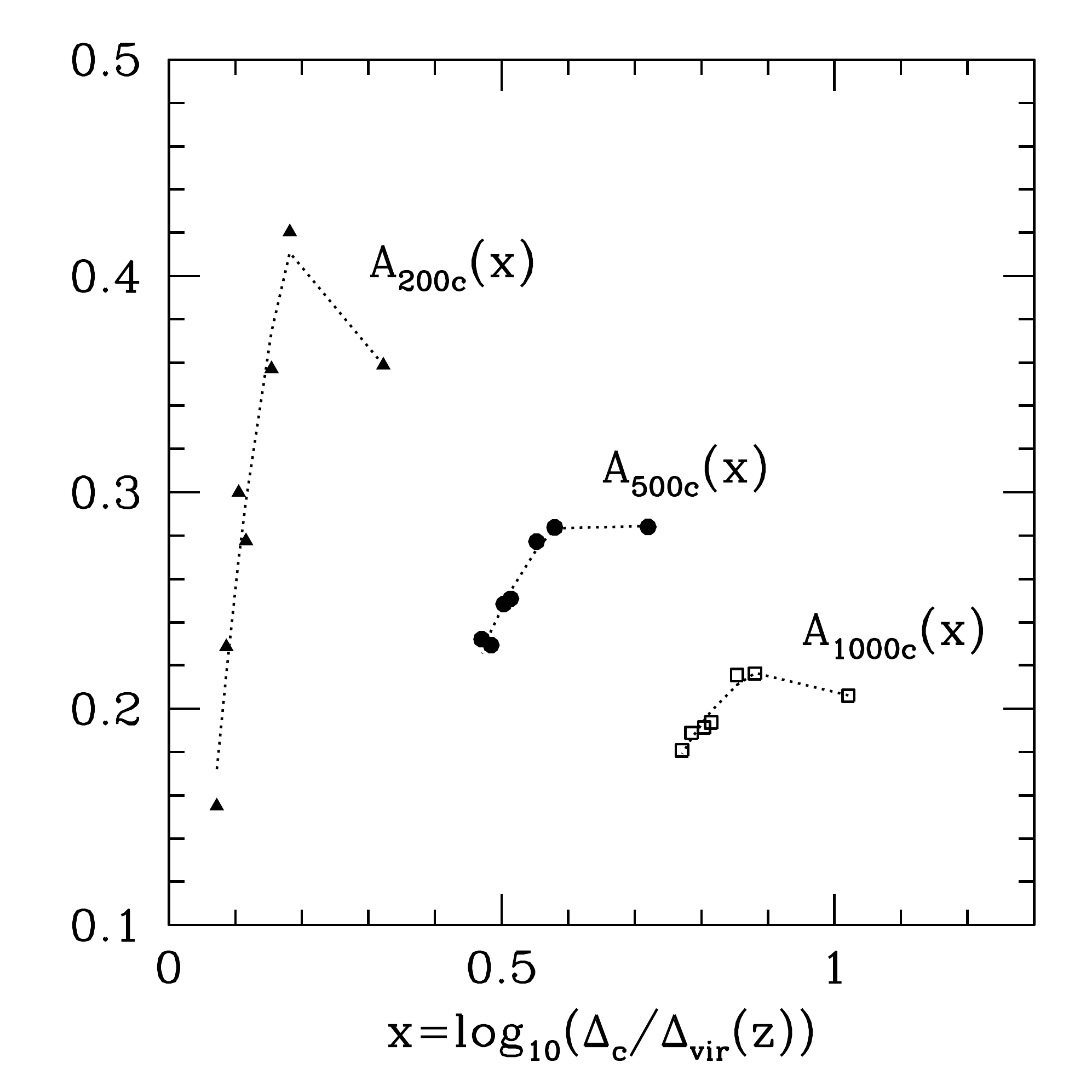}
\includegraphics[width=.4\textwidth]{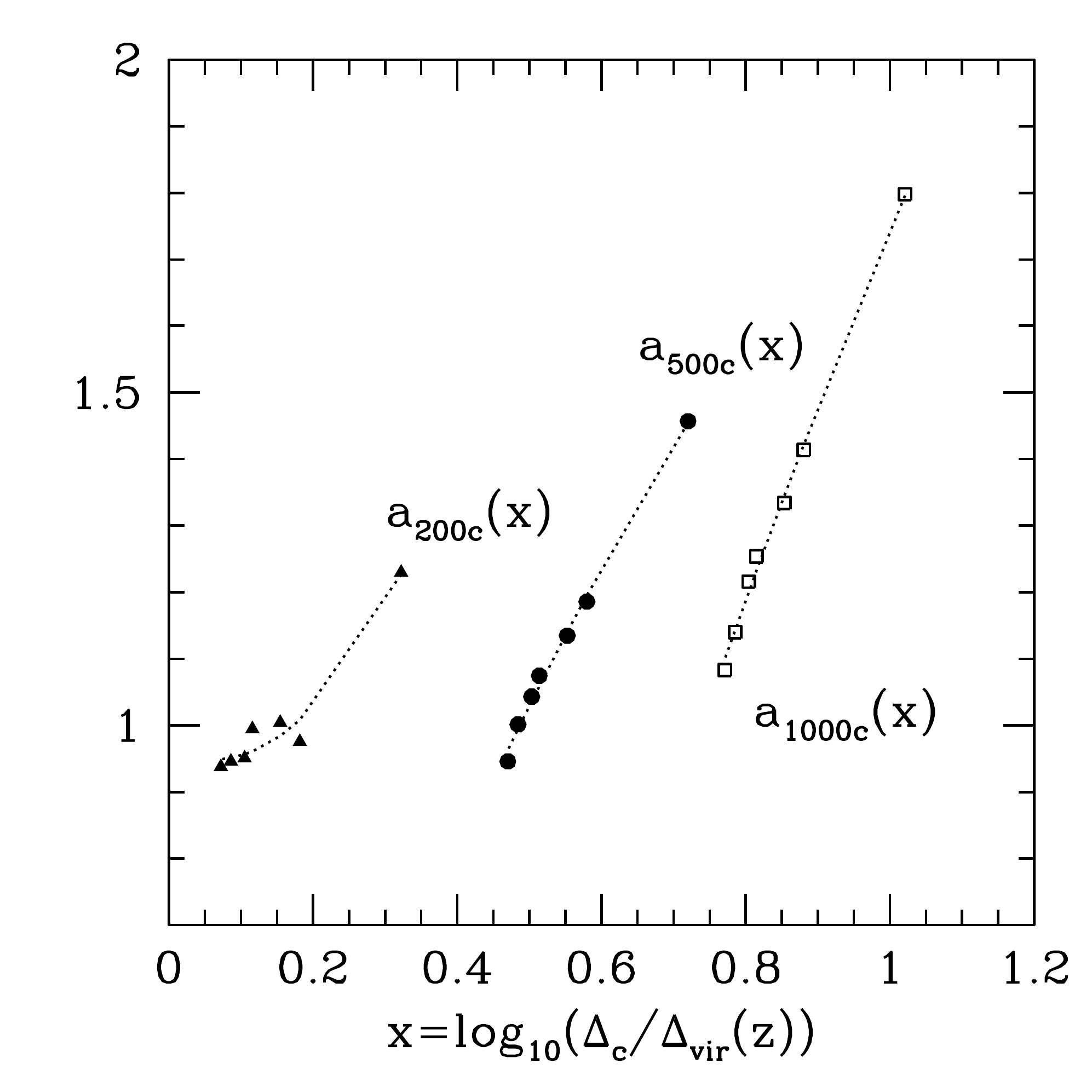}
\includegraphics[width=.4\textwidth]{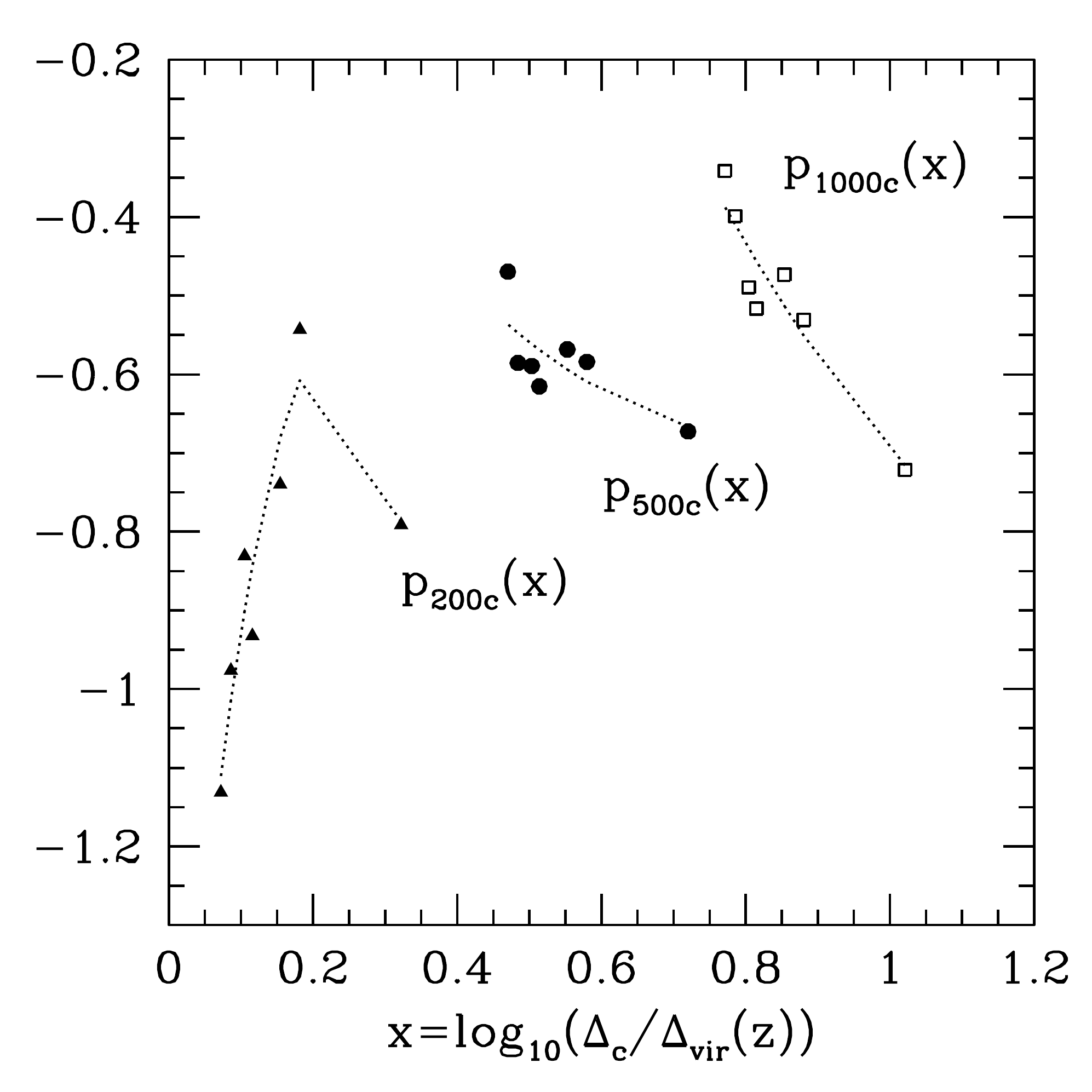}
\caption{Comparison of the ST best-fit parameters dependence on $x=\log_{10}(\Delta_c/\Delta_{vir}(z))$ and the best-fit quadratic functions for $A_{\Delta_c}$ (top left panel), $a_{\Delta_c}$ (top right panel) and $p_{\Delta_c}$ (bottom panel) for $\Delta_c=200$, $\Delta_c=500$ and $1000$, respectively.}
\label{st_coef_trend}
\end{figure}

In order to extrapolate the mass functions at any given redshift, we follow the approach of \citet{Despali2016} and parametrise the redshift dependence of the ST coefficients in terms of the variable $x=\log_{10}(\Delta/\Delta_{vir}(z))$, where $\Delta_{vir}(z)$ is the virial overdensity as given by the formula derived in \citet{Bryan1998}. We find that the redshift variation of the best-fit ST coefficients can be described to very good approximation by a quadratic fit as function of $x$ given by:

\begin{equation}\label{coefm200c}
\begin{cases}
\quad A_{200c}(x)=-10.2185312\,x^2+4.78051093\,x-0.1206716\\
\quad a_{200c}(x)=4.07275047\,x^2-0.49618532\,x+0.96372361\\
\quad p_{200c}(x)=-23.48761585\,x^2+10.5651697\,x-1.752599071
\end{cases}
\end{equation}

\begin{equation}\label{coefm500c}
\begin{cases}
\quad A_{500c}(x)=-2.08511667\,x^2+2.71726345\,x-0.59113241\\
\quad a_{500c}(x)=-1.0788725\,x^2+3.25302957\,x-0.32810261\\
\quad p_{500c}(x)=1.04288295\,x^2-1.76269479\,x+0.06162189
\end{cases}
\end{equation}
and
\begin{equation}\label{coefm1000c}
\begin{cases}
\quad  A_{1000c}(x)=-1.65696205\,x^2+3.07836133\,x-1.20944538 \\
\quad  a_{1000c}(x)=-1.18612053\,x^2+4.91186256\,x-1.98337952\\
\quad  p_{1000c}(x)= 1.33135179\,x^2-3.7042898\,x+1.67853762\\
\end{cases}
\end{equation}
which we plot in Fig.~\ref{st_coef_trend} against the best-fit values quoted in Table~\ref{tabA1}. 

\begin{figure}
\centering
\includegraphics[width=.4\textwidth]{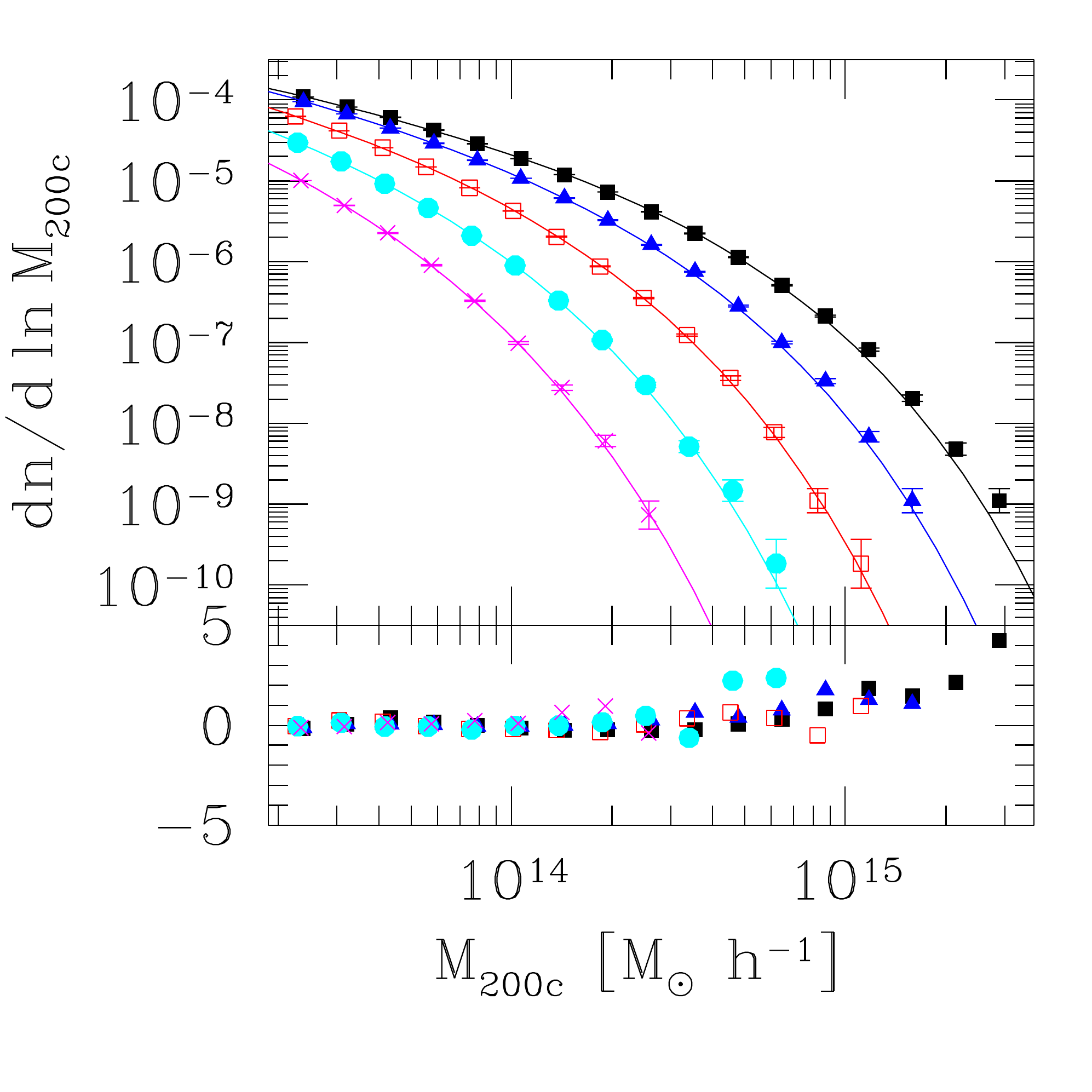}
\includegraphics[width=.4\textwidth]{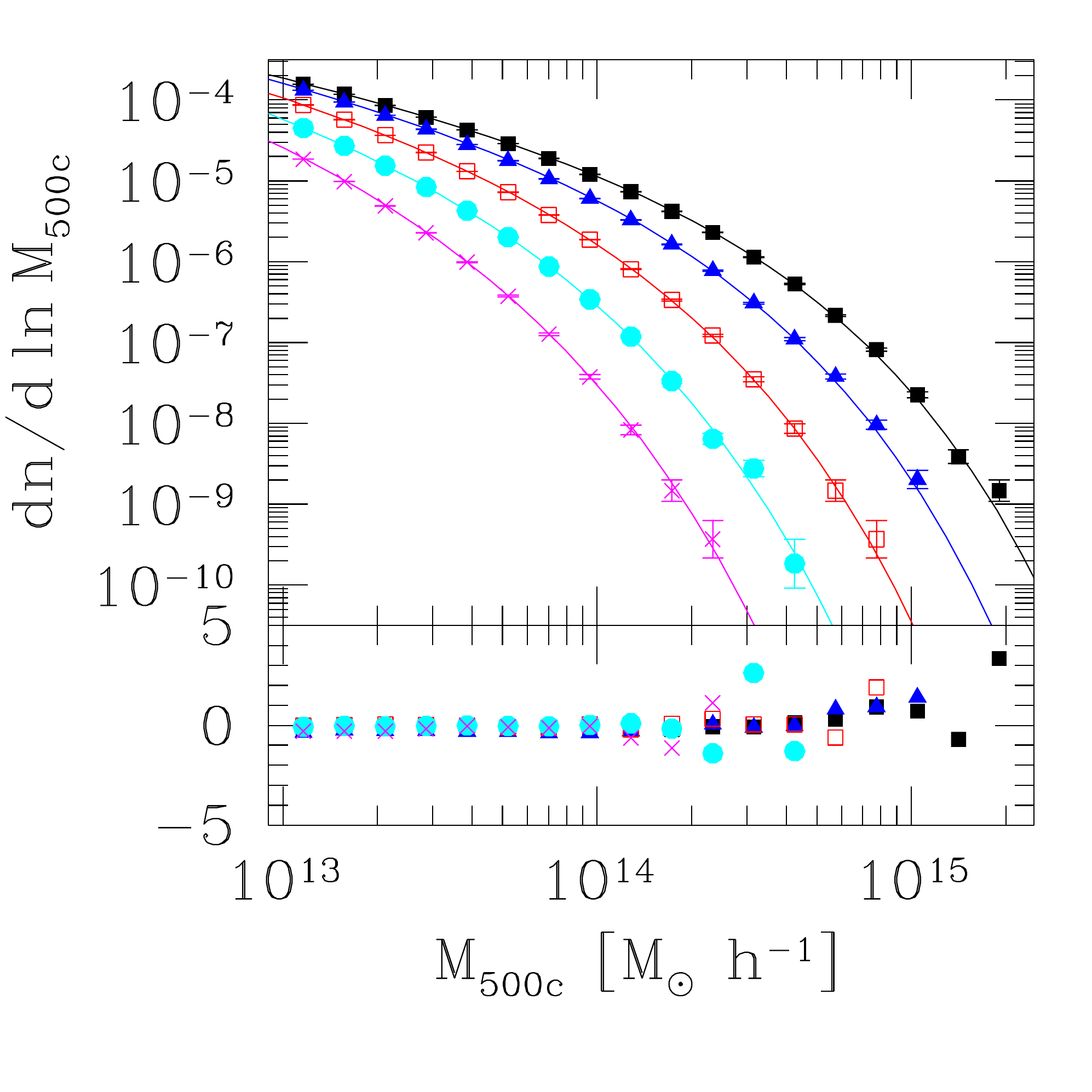}
\includegraphics[width=.4\textwidth]{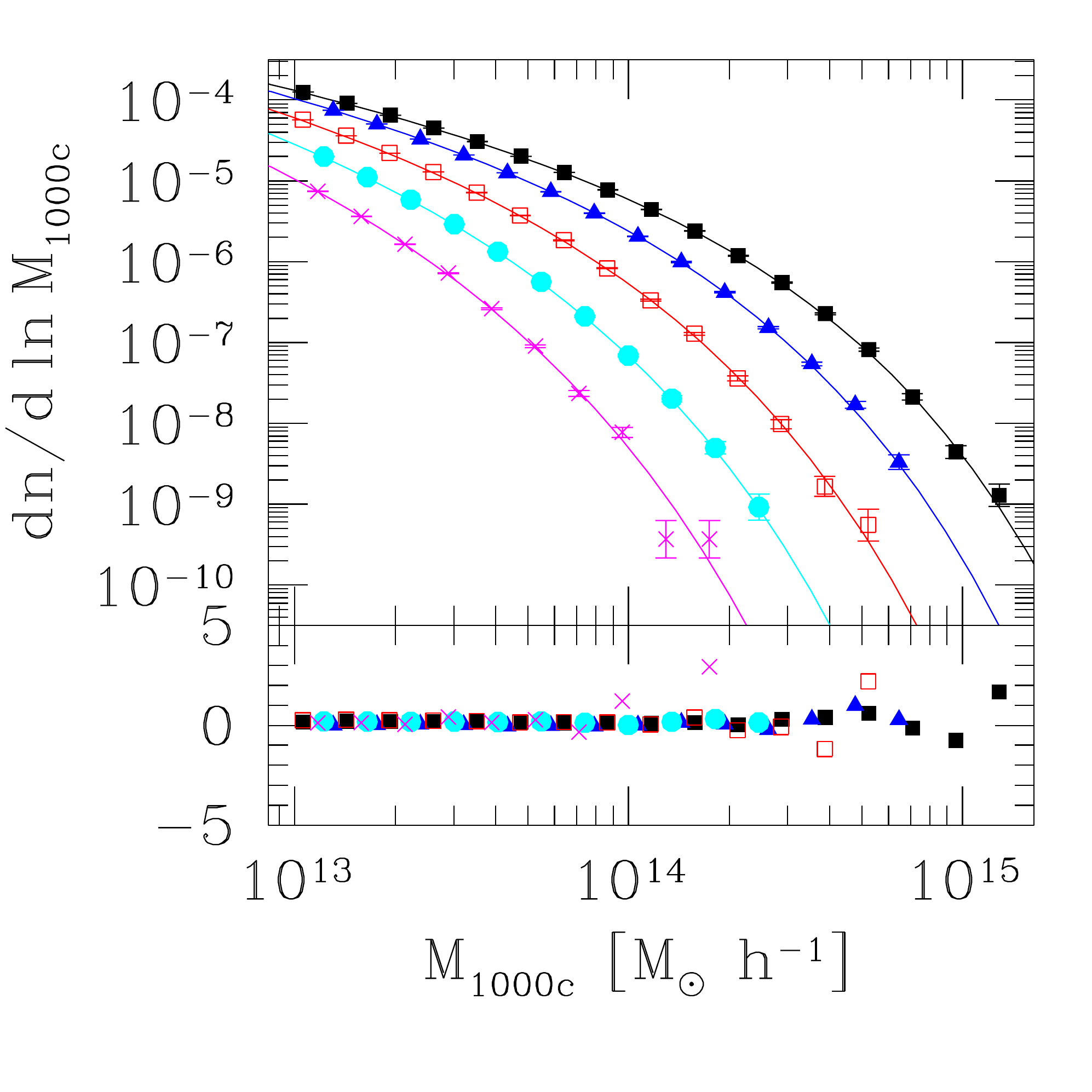}
\caption{Halo mass function from the RayGalGroupSims simulation for SOD halos with mass  ${\rm M}_{200c}$ (top left panel) ${\rm M}_{500c}$ (top right panel) and ${\rm M}_{1000c}$ (bottom panel) at $z=0,0.5,1,1.5$ and $2$ (top to bottom) respectively. The solid lines are the ST-RayGal mass functions with coefficients given by Eq.~(\ref{coefm200c}), Eq.~(\ref{coefm500c}) and Eq.~(\ref{coefm1000c}) respectively. The logarithmic residual is shown in the bottom panel: as we can see deviations are within $5\%$ level across the entire mass range.}
\label{hmf_app}
\end{figure}

In Fig.~\ref{hmf_app} we plot the ST mass functions for ${\rm M}_{200c}$, ${\rm M}_{500c}$ and ${\rm M}_{1000c}$ with coefficients given by Eq.~(\ref{coefm200c}), Eq.~(\ref{coefm500c}) and Eq.~(\ref{coefm1000c}) against the N-body mass function estimates, to which we have referred as the ST-RayGal mass functions. As we can see logarithmic deviations with respect to the numerical results are still within $5\%$ level.

\begin{figure}[th]
\centering
\includegraphics[width=.4\textwidth]{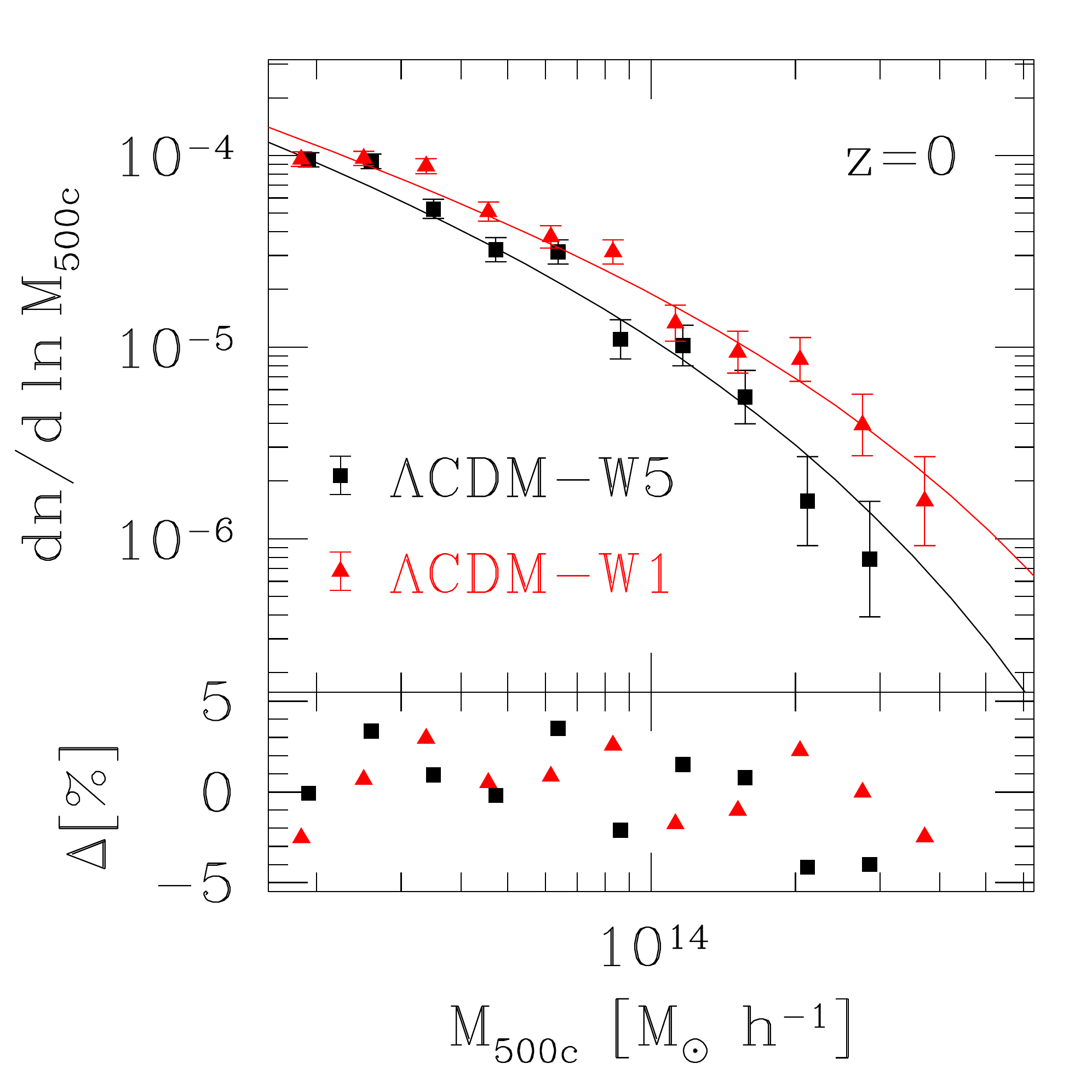}
\includegraphics[width=.4\textwidth]{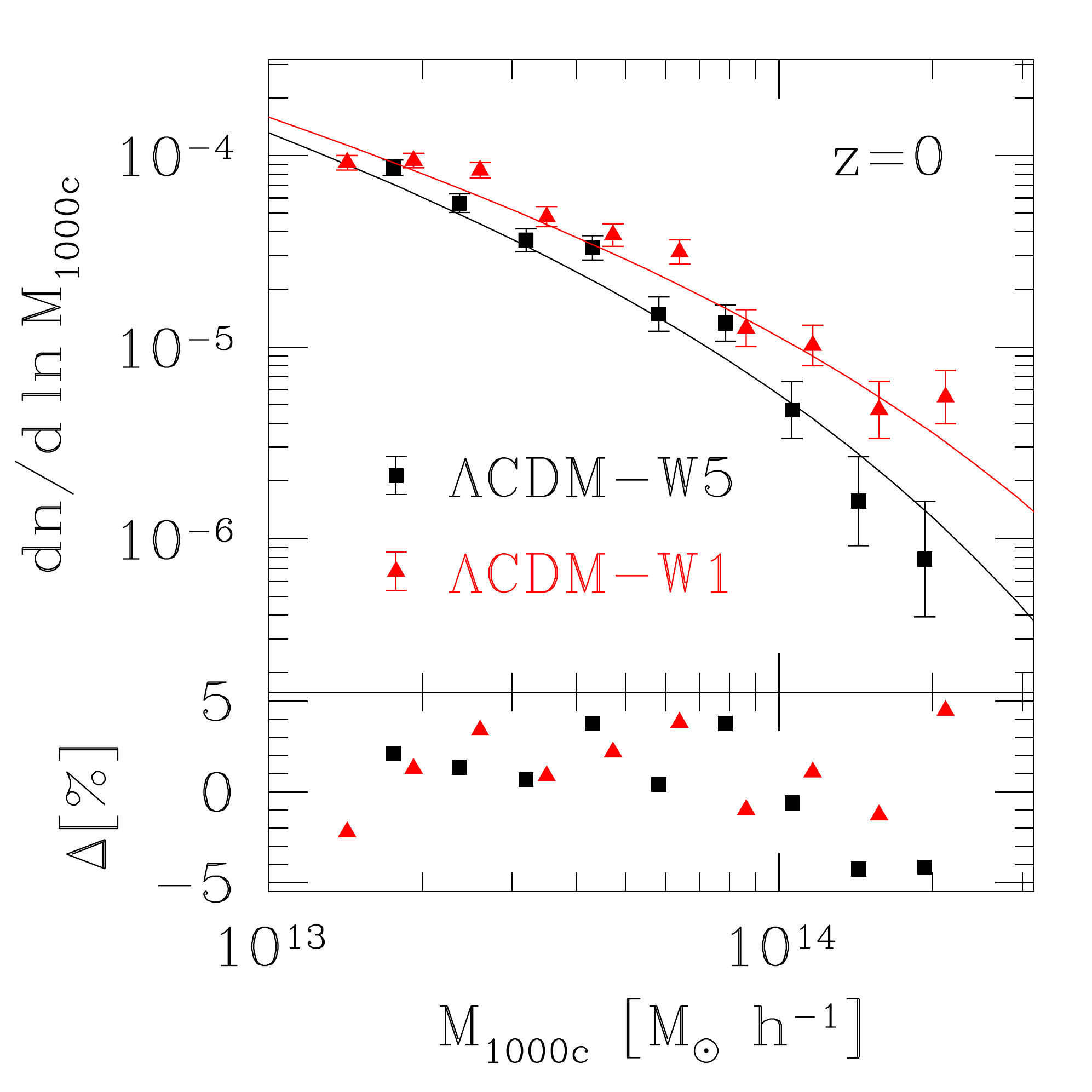}
\caption{SOD halo mass functions from the $\Lambda$CDM-W5 (black squares) and $\Lambda$CDM-W1 (red triangles) with mass ${\rm M}_{500c}$ (left panel) and ${\rm M}_{1000c}$ (right panel) respectively against the ST-RayGal predictions.}
\label{mf_wmap15}
\end{figure}

We find that the ST-RayGal mass function formulae can also reproduce the SOD mass functions for ${\rm M}_{500c}$ and ${\rm M}_{1000c}$ from N-body simulations with different cosmological parameter values. In particular, we have used halo catalogs at $z=0$ from simulations of $162$ h$^{-1}$ Mpc box-length and $512^3$ particles of two flat $\Lambda$CDM models: $\Lambda$CDM-W1 with $\Omega_m=0.29$, $\sigma_8=0.90$, $\Omega_b=0.047$ and $n_s=0.990$; $\Lambda$CDM-W5 with $\Omega_m=0.26$, $\sigma_8=0.79$, $\Omega_b=0.044$ and $n_s=0.963$. As shown in Fig~\ref{mf_wmap15}, the logarithmic differences between the ST-RayGal mass function and the numerical estimates from the $\Lambda$CDM-W5 and $\Lambda$CDM-W1 catalogs respectively are within the $5\%$ level. Using the same halo catalogs we estimate the average sparsity at $z=0$. In the case of the $\Lambda$CDM-W5 simulation we find $\langle s_{500,1000} \rangle=1.41$, while in the $\Lambda$CDM-W1 case we find $\langle s_{500,1000} \rangle=1.36$. These values are consistent to within a few percent with the average sparsity prediction inferred by solving Eq.~(\ref{sparpred}) with the ST-RayGal mass functions. In particular, we obtain $\langle s^{\rm th}_{500,1000} \rangle=1.43$ for the $\Lambda$CDM-W5 model and $\langle s^{\rm th}_{500,1000} \rangle=1.39$ for the $\Lambda$CDM-W1 cosmology. 


\end{document}